%% file: main.tex
\documentclass[conference,compsoc]{IEEEtran}

\newif\ifArxivVersion

\ArxivVersiontrue %

\NewDocumentCommand{\ArxivOnly}{+m}{%
  \ifArxivVersion#1\fi%
}

\NewDocumentCommand{\SubmissionOnly}{+m}{%
  \ifArxivVersion\else#1\fi%
}

\usepackage[]{hyperref}
\ArxivOnly{\hypersetup{hidelinks}}

\usepackage[table]{xcolor}
\usepackage{tikz}
\usepackage{amsmath}
\usepackage{graphicx}
\usepackage{amsthm}
\usepackage{comment}

\theoremstyle{definition}
\newtheorem{definition}{Definition}
\theoremstyle{plain}
\newtheorem{theorem}{Theorem}

\input{macros}

\begin{document}

\title{\Large \bf A Star is Reborn: The PHP Interpreter Rewritten in Safe Rust}

\SubmissionOnly{
    \author{\IEEEauthorblockN{Anonymous Author(s)}}
}

\ArxivOnly{
    \author{
        \IEEEauthorblockN{
            Tianyu Li \quad
            Bo Wang \quad
            Prateek Saxena \quad
            Umang Mathur
        }
        \IEEEauthorblockA{
            National University of Singapore \\
            tianyuli@u.nus.edu,
            bo-wang@nus.edu.sg,
            dcsprs@nus.edu.sg,
            umathur@nus.edu.sg
        }
    }
}

\maketitle

\input{sections/00-abstract}
\input{sections/01-introduction}

\input{sections/02-background}
\input{sections/03-overview}

\input{sections/04-memory-model}
\input{sections/05-refinement-verification}

\input{sections/06-implementation}
\input{sections/07-Evaluation}
\input{sections/07-findings-security}
\input{sections/08-findings-performance}
\input{sections/09-findings-functionality}
\input{sections/10-case-study}

\input{sections/11-related}
\input{sections/12-conclusion}

\bibliographystyle{plainurl}
\bibliography{references}

\input{sections/99-appendix}

\end{document}

%% file: macros.tex
\usepackage[table]{xcolor}
\usepackage{multirow}
\usepackage{xcolor}
\usepackage{soul}          %
\makeatletter
\g@addto@macro\UrlBreaks{\do\-\do\.\do\/\do\_\do\?\do\&\do\=\do\%\do\#\do\:}
\makeatother
\usepackage{amssymb}
\usepackage{booktabs}
\usepackage{enumitem}
\usepackage{amsmath}
\usepackage{booktabs}
\usepackage{stmaryrd}
\usepackage{tikz}
\usetikzlibrary{arrows.meta}
\usepackage{graphicx}
\usepackage{tikz}
\usetikzlibrary{arrows.meta,positioning,fit,backgrounds}

\usepackage{array}
\usepackage{listings}      %
\usepackage[T1]{fontenc}
\usepackage[scaled=0.82]{beramono}

\definecolor{dnumbg}{RGB}{212,237,218}   %
\definecolor{dnumfg}{RGB}{22,101,52}      %
\definecolor{todobg}{RGB}{207,226,255}    %
\definecolor{todofg}{RGB}{0, 110, 255}      %

\usepackage{xcolor}

\def\commentmode{2}

\ifnum\commentmode=1
    \newcommand{\prateek}[1]{\textcolor{red}{Prateek: #1}}
    \newcommand{\tianyu}[1]{\textcolor{blue}{Tianyu: #1}}
    \newcommand{\bo}[1]{\textcolor{purple}{Bo: #1}}
    \newcommand{\umang}[1]{\textcolor{green!60!black}{Umang: #1}}
    
\else
    \newcommand{\prateek}[1]{}
    \newcommand{\tianyu}[1]{}
    \newcommand{\bo}[1]{}
    \newcommand{\umang}[1]{}
    
\fi

\usepackage{booktabs,array,threeparttable,makecell}

\usepackage[most]{tcolorbox}
\newcommand{\code}[1]{\texttt{#1}}

\lstdefinelanguage{Rust}{
  morekeywords={enum,struct,impl,fn,let,mut,match,pub,use,mod,type,trait,
                self,Self,ref,move,where,for,in,if,else,return,unsafe,dyn,as},
  sensitive=true,
  morecomment=[l]{//},
  morecomment=[s]{/*}{*/},
  morestring=[b]",
}

\definecolor{planfg}{RGB}{146,64,14}      %
\definecolor{planbg}{RGB}{254,243,199}    %
\definecolor{planrule}{RGB}{217,119,6}    %

\definecolor{lstbg}{gray}{0.955}     %
\definecolor{lstrule}{gray}{0.45}    %
\definecolor{lstkw}{RGB}{0,0,160}
\definecolor{lstcm}{RGB}{95,95,95}
\definecolor{lstst}{RGB}{150,30,30}

\lstdefinelanguage{PHP5}{
  morekeywords={abstract,and,array,as,break,case,catch,class,clone,const,
                continue,declare,default,do,echo,else,elseif,empty,extends,
                final,for,foreach,function,global,if,implements,include,
                include_once,instanceof,interface,isset,list,new,or,print,
                private,protected,public,require,require_once,return,static,
                switch,throw,try,unset,use,var,while,xor,true,false,null},
  sensitive=true,
  morecomment=[l]{//},
  morecomment=[s]{/*}{*/},
  morestring=[b]",
  morestring=[b]',
}

\usepackage{xcolor}
\usepackage{listings}

\lstdefinestyle{phpstyle}{
    language=PHP,
    basicstyle=\ttfamily\footnotesize,
    keywordstyle=\color{blue!65!black},
    commentstyle=\color{gray!85!black}\itshape,
    stringstyle=\color{red!55!black},
    numbers=left,
    numberstyle=\tiny\color{gray},
    numbersep=6pt,
    breaklines=true,
    breakatwhitespace=true,
    columns=fullflexible,
    keepspaces=true,
    showstringspaces=false,
    frame=single,
    framesep=4pt,
    xleftmargin=8pt,
    captionpos=b
}

\lstdefinestyle{paperlst}{
  basicstyle=\ttfamily\footnotesize,
  keywordstyle=\color{lstkw}\bfseries,
  commentstyle=\color{lstcm}\itshape,
  stringstyle=\color{lstst},
  columns=fixed, keepspaces=true,       %
  showstringspaces=false,
  breaklines=true, breakautoindent=true, breakindent=10pt,
  postbreak=\mbox{\textcolor{lstrule}{$\hookrightarrow$}\space},
  frame=single, framerule=0.7pt, framesep=5pt,
  rulecolor=\color{lstrule},
  backgroundcolor=\color{lstbg},
  xleftmargin=2pt, xrightmargin=2pt,
  aboveskip=8pt, belowskip=2pt,
  captionpos=b, abovecaptionskip=5pt,
}

%% file: sections/00-abstract.tex
\begin{abstract}

PHP powers most of the web,
yet its C interpreter has had many memory-safety vulnerabilities.
Reimplementing the interpreter in safe Rust offers a promising way to eliminate these vulnerabilities.
However, eliminating \emph{native} memory violations alone does not necessarily prevent unintended accesses to the \emph{logical} memory of the translated program.

To prevent such failures, we introduce \emph{logical memory safety},
which confines the program operations to their permitted memory.
In this paper, we formalize this property for PHP through an abstract PHP machine model that captures how the PHP interpreter manages program values while abstracting away functionality details.
Guided by this model, we reimplement the PHP~5.0.0 engine in safe Rust,
and formally prove that, for a core fragment of PHP, our Rust translation \emph{refines} the abstract PHP machine, thereby establishing logical memory safety for this fragment.

Our translation safely handles all $173$ memory vulnerabilities we collected that affect PHP 5.0.0 prior to the rewrite. The Rust version matches the C implementation on all functionality tests, and produces identical outputs for the real-world PHP applications we evaluate.
To our knowledge, this is the first principled safe Rust port of a core fragment of the PHP interpreter that combines formal guarantees of logical memory safety with functional compatibility.
The improved security comes at a modest average runtime overhead of only \emph{15\%} compared to C.

\end{abstract}

%% file: sections/01-introduction.tex
\section{Introduction}
\label{sec:intro}

PHP is a widely used language. Recent reports estimate that more than 60\% of websites use PHP{\footnote{Among sites whose server-side language can be identified}}~\cite{w3techs-php}.
Its interpreter engine is predominantly written in C,
enabling low-level control over runtime data structures for performance and memory efficiency~\cite{zend-mm, php-memory,php-gc}.
However, C does not enforce memory safety, leaving the PHP interpreter vulnerable to numerous memory-safety bugs throughout its history~\cite{jiang2025flowfusion, php-memory-hardening}. A prime example is the recent PHP2Shell vulnerability~\cite{php2shell}, which enables unauthenticated code execution.
Patches for memory vulnerabilities can take months to be deployed,
creating an exploitable window between vulnerability discovery and deployment of a fix~\cite{demir2021insecure}.

One promising solution is to reimplement the interpreters in safe Rust, which enforces memory safety through a combination of compile-time guarantees and runtime checks~\cite{rustlan}.
At the same time, Rust retains low-level control over memory allocation, data representation, and resource lifetimes, making it well suited for implementing performance-sensitive language runtimes.
Recent advances in coding agents are also making large-scale C-to-safe-Rust translation increasingly practical~\cite{actor,wang2026reboot}.

\begin{figure}[t]
    \centering
    \includegraphics[width=\linewidth]{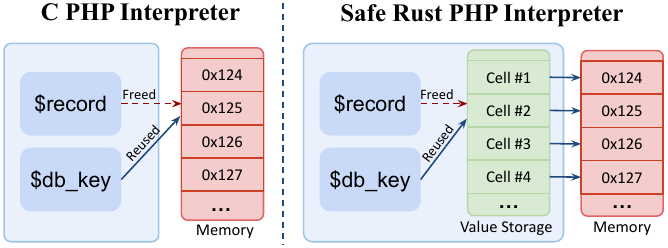}
    \caption{The conceptual demonstration of how a C use-after-free bug survives as an unintended PHP state access in safe Rust.
    }
    \label{fig:example}
\end{figure}

It is tempting to believe that if C is rewritten into safe Rust, i.e., with no {\tt unsafe} code, then all memory errors have been corrected. This is not true. We find that when translating a real-world interpreter such as PHP, implementation-level memory (\emph{native memory}) safety alone does not always rule out unintended accesses to the objects (\emph{logical memory}) of the interpreted program.

As shown in Figure~\ref{fig:example}, the C PHP engine has a memory error wherein a premature reclamation leaves the variable holding a dangling pointer. When the rewritten Rust stores PHP values in an indexed container, the original memory error in C can turn into a logical error: a stale index pointing in the container.
If the interpreter subsequently reuses that index for another value containing sensitive data,
the operation holding the stale index may then access a sensitive value in the PHP logical memory state,
even though the original PHP script has no legitimate way to access it.
Safe Rust removes the original memory-safety violation at the level of raw memory addresses in C, but the unintended access to sensitive PHP state may persist logically.

How to translate the C program to safe Rust that eliminates both native memory errors and logical memory violations?
The latter guarantee depends on the \emph{intent} of the program, which determines what value the execution may legitimately access and how values may be shared.

In this paper, we investigate this challenge for language interpreters,
using PHP as a concrete example. The PHP interpreter is written in C. We first translate it into safe Rust, thereby ensuring no native memory errors in the target language.
The PHP interpreter also manages logical memory state through C data structures, using pointers, reference counting, and explicit memory management.
We formalize the relevant PHP memory semantics in an \emph{abstract PHP machine model}, based on the PHP language specification~\cite{php-memory-spec}.

We then formalize \emph{logical memory safety}, which intuitively prevents these logical memory violations in the abstract PHP machine model semantics.
The safety property captures a basic requirement that an interpreter should guarantee:
Interpreter operations should not access or modify program states that are not permitted by their operands.

\vskip 0.05in
\noindent\textbf{Implementation.}
Rewriting a large interpreter is costly, even with
coding agents.
In this work, we aim to demonstrate our core approach and provide evidence that its principles are practically realizable.
We therefore work with PHP~5.0.0,
a prior version of the PHP interpreter. PHP-5.0.0 has 37K lines of code in the engine, and is thus within our computational budget to rewrite into fully safe Rust. Nonetheless, this version still retains the core PHP memory semantics intended of the latest versions of the interpreter.

We rewrite the interpreter following the abstract PHP machine model.
We then formally prove that a core part of the interpreter engine in our concrete Rust implementation {\em refines}~\cite{abadi1991refinement} the abstract PHP model and mechanize the proof through Verus~\cite{lattuada2023verus}.

For experimental confirmation,
we collected $173$ historically known memory-related vulnerabilities that affect the C version of our original PHP interpreter.
Our safe Rust behaves safely on all these vulnerabilities with no native memory-safety violations or logical memory-safety violations.

Our rewrite produces byte-identical outputs to C PHP on 8 evaluated applications, including WordPress~\cite{wordpress2.0.11}, and on all 1,143 standalone PHP tests.
It incurs an average of only 15\% runtime overhead on the PHP performance benchmark.

\begin{tcolorbox}[
  colback=blue!4,
  colframe=blue!35!black,
  boxrule=0.5pt,
  arc=1pt,
  left=6pt,
  right=6pt,
  top=5pt,
  bottom=5pt
]

\noindent\textbf{Main Contribution.}
We demonstrate how to establish both \emph {native} and \emph{logical} memory safety
for C-to-Rust translation.
We instantiate this approach for PHP using an abstract PHP machine, and demonstrate its feasibility through a substantial interpreter rewrite with functional compatibility and modest runtime overhead.
\end{tcolorbox}

%% file: sections/02-background.tex
\section{Motivation and Overview}
\label{sec:background}

To see how safe Rust prevents native memory violations and what security risks may remain, we examine a bug in the C PHP interpreter and its potentially harmful behavior after a safe Rust rewrite.

\subsection{PHP Execution and Memory Semantics}
\label{sec:php-background}

\noindent\textbf{PHP request execution.}
PHP is commonly deployed as a server-side interpreter for web applications.
Upon receiving an HTTP request, the server invokes the PHP interpreter to compile the application script into opcode arrays (\emph{op arrays}), which contain instructions (\emph{opcodes}) executed by the Zend virtual machine~\cite{zend-mm}.
It then executes the application script with inputs from the requests.
During execution, the interpreter maintains PHP program state, including variables, values, and objects created or modified by the script.

\noindent\textbf{PHP memory semantics.}
The PHP language specification~\cite{php-memory-spec} conceptually defines a memory model that PHP script execution should follow.
PHP distinguishes variables from the values they hold and supports several forms of sharing.
For ordinary value assignment, \texttt{\$x = \$y}, the variables remain independent: subsequent assignments to \texttt{\$x} do not change the value of \texttt{\$y}.
In contrast, reference assignment, \texttt{\$x =\& \$y}, makes the two variables aliases, so updating either variable affects the value observed through the other.
PHP objects have different sharing semantics from variables.
For example, after \texttt{\$x = \$obj;}, both variables refer to the same object, and modifying an object property through \texttt{\$x} is observable through \texttt{\$obj}.

\subsection{A Memory Bug in the C PHP Engine}
\label{sec:c-engine-bug}
The Zend Engine~\cite{zend-mm} implements PHP's memory semantics in C, using pointers, reference counting, and explicit memory management mechanisms.
Errors in managing these structures can leave PHP values referring to reclaimed memory, resulting in memory-safety violations.
The example below demonstrates one such error.

Listing~\ref{lst:php-motivating} shows a simplified PHP web application that processes user requests.
The application parses a user-submitted record, stores the record in a database, and returns it to the user.
Although the script behaves as intended under correct PHP semantics, it triggers a use-after-free bug in the C PHP~5.0.0 interpreter.

The bug is triggered at line~4,
where the script uses \texttt{unserialize} to parse the user input.
On a crafted input that contains back-references\footnote{
A legitimate feature for encoding shared and recursive values.
The crafted input makes an element refer back to itself;
in PHP~5.0.0 the handler then frees that element's own value while the array
is still referring to it.
},
\texttt{unserialize} returns a PHP array whose \texttt{user} field holds a value that has been \emph{reclaimed} by the interpreter.
Subsequent accesses to this field dereference a dangling pointer, where AddressSanitizer reports a use-after-free.
Such kinds of memory errors may enable information leakage, data corruption, or control-flow hijacking~\cite{roemer2012return, pincus2004beyond}.

\noindent\textbf{Reimplement in Safe Rust.}
Reimplementing the interpreter in safe Rust prevents this use-after-free at the native memory level.
Rust's ownership and borrowing rules prevent safe code from dereferencing a reference after its storage has been freed by construction,
without requiring programmers to check the validity of individual memory accesses.
Therefore, invalid accesses to native memory will be eliminated.

However, the security harm from the native memory bug may persist via a different path without a native memory violation.
Therefore, eliminating native memory violations does not necessarily eliminate all the security harm it can enable.
The example below shows how the data-leakage enabled by the use-after-free bug above can be preserved in a fully memory-safe translation without violating Rust's native memory safety.

\subsection{An Analogous Bug Preserved in Safe Rust}
\label{sec:motivation}

\begin{figure}[t]
\begin{lstlisting}[
    style=phpstyle,
    caption={A benign PHP server script that triggers a use-after-free bug in C PHP~5.0.0. When interpreted by a vulnerable safe Rust PHP engine, it may expose the secret key "S3CRET-DB-KEY" which the script does not intend to disclose.},
    label={lst:php-motivating}
]
<?php
function save_to_database($record, $key){...}

$record = unserialize($_COOKIE['sess']); // bug trigger

$DB_KEY = "S3CRET-DB-KEY";
save_to_database($record, $DB_KEY); // store in database

echo $record['user']; // reflect to user
\end{lstlisting}
\end{figure}

Consider a straightforward translation strategy of the PHP interpreter into safe Rust.
The translated engine stores PHP values in an indexed container and uses integer indices for accessing them.
The engine manages freed slots by a last-in, first-out (LIFO) free list, so the next allocation reuses the most recently released slot.

Listing~\ref{lst:php-motivating} shows how an incorrect \texttt{unserialize} leaves the \texttt{user} field holding a stale integer
index:
the slot it accesses has already been reclaimed.
In the Rust translation, the reclaimed slot goes on the free list,
so when the database key \texttt{"S3CRET-DB-KEY"} is subsequently allocated,
the engine reuses that slot.
The \texttt{user} field now refers to the database key  through the stale index.
When the application responds with the record, the interpreter reads the key
and discloses it to the requester.

All accesses in the translated Rust engine remain natively memory-safe: there is no raw pointer usage, the access to the container is within bounds, and the reused slot contains a valid PHP value.
Nevertheless, the interpreter exposes a secret that the application did not intend to disclose.
Such errors may enable data-only exploits~\cite{chen2005noncontrol,hu2016dop},
such as leaking sensitive information or corrupting credential values in the logical memory of the interpreter.

In summary, the original native memory error in the C PHP interpreter did not disappear in the Rust rewrite---it merely morphed into what we call a logical memory error.

\subsection{Landscape of PHP Memory Bugs}
\label{sec:bug-landscape}

\begin{table}[t]
  \centering
  \footnotesize
  \caption{Distribution of historical PHP memory bugs
  by defect class and affected component.
  }
  \label{tab:sec-bench}
  \begin{tabular}{@{}l r r r@{}}
    \toprule
    \textbf{Defect class}
      & \textbf{Engine} & \textbf{Library} & \textbf{Total} \\
    \midrule

    \multicolumn{4}{@{}l}{\emph{C memory corruptions}} \\
    Use-after-free                 & 33 & 22 & 55 \\
    Out-of-bounds read             & 5  & 17 & 22 \\
    NULL-pointer dereference       & 8  & 6  & 14 \\
    Integer overflow/wraparound    & 2  & 11 & 13 \\
    Out-of-bounds write            & 1  & 11 & 12 \\
    Type confusion                 & 5  & 6  & 11 \\
    Stack-address escape           & 4  & 1  & 5  \\
    Invalid free                   & 3  & 1  & 4  \\
    Double-free                    & 1  & 1  & 2  \\
    Uninitialized resource         & -- & 1  & 1  \\
    Uninitialized pointer          & 3  & -- & 3  \\
    \cmidrule(l){2-4}
    \emph{Subtotal} & \emph{65} & \emph{77}
                       & \emph{142} \\

    \addlinespace[3pt]
    \multicolumn{4}{@{}l}{\emph{PHP memory semantic errors}} \\
    Extra reference binding        & 9  & 2  & 11 \\
    Lost reference binding         & 2  & 6  & 8  \\
    Improper resource lifetime     & 5  & -- & 5  \\
    Missing memory release         & 5  & -- & 5  \\
    Incorrect reference count      & 1  & -- & 1  \\
    Premature release              & 1  & -- & 1  \\
    \cmidrule(l){2-4}
    \emph{Subtotal} & \emph{23} & \emph{8} & \emph{31} \\

    \midrule
    \textbf{Total} & \textbf{88} & \textbf{85}
                      & \textbf{173} \\
    \bottomrule

  \end{tabular}
\end{table}

To examine the prevalence of these underlying error classes,
we collect 173 historical memory vulnerabilities affecting PHP~5.0.0.
Table~\ref{tab:sec-bench} summarizes their distribution
across categories and locations.
142 of them are native memory bugs in the C PHP,
and the remaining 31 are memory-management vulnerabilities, such as refcounting errors, value-binding errors, and so on.
These 31 violate the PHP semantics within our targeted protection scope, but they do not result in memory corruption.
Use-after-free and out-of-bounds accesses account for a substantial portion of the bugs.
About half of the bugs we collected are rooted in the interpreter engine; the remaining half are in surrounding built-in libraries or extensions.
Appendix~\ref{app:sec-bench} explains the classification and how we collect these bugs.

To assess whether these vulnerabilities could persist in a
safe Rust implementation, we examine their underlying
failure patterns.
We find that at least 56 involve logical memory access violations that safe Rust alone does not prevent.
Section~\ref{sec:eval-security} presents the detailed analysis.

\subsection{Threat Model and Security Goals}
\label{sec:threat-model}

\noindent\textbf{Attacker capabilities.}
We consider PHP applications that execute on a potentially vulnerable interpreter and process both public and sensitive data.
We assume the application does not intentionally disclose sensitive data or perform unauthorized state modifications.
The attacker can invoke application functionality through HTTP requests and observe the resulting responses, but cannot directly modify the interpreter or the application's source code.

\noindent\textbf{(a) Native memory violations.}
In a memory-unsafe interpreter,
bugs such as out-of-bounds accesses and use-after-free may compromise the native memory of the interpreter process.
Our safe Rust rewrite prevents these native memory violations by construction.

\noindent\textbf{(b) Logical memory violations.}
A native memory-safe interpreter may still access or modify values in the logical memory state that are not permitted.
We aim to prevent these logical memory violations beyond native memory safety.

%% file: sections/03-overview.tex
\section{Approach Overview}
\label{sec:overview}

Our translation goal is to ensure both native and logical memory safety while preserving the interpreter's functionality and performance.
The key challenge is balancing the safety guarantees we want with the practicality of verifying the final implementation.

\begin{figure*}
    \centering
    \includegraphics[width=\textwidth]{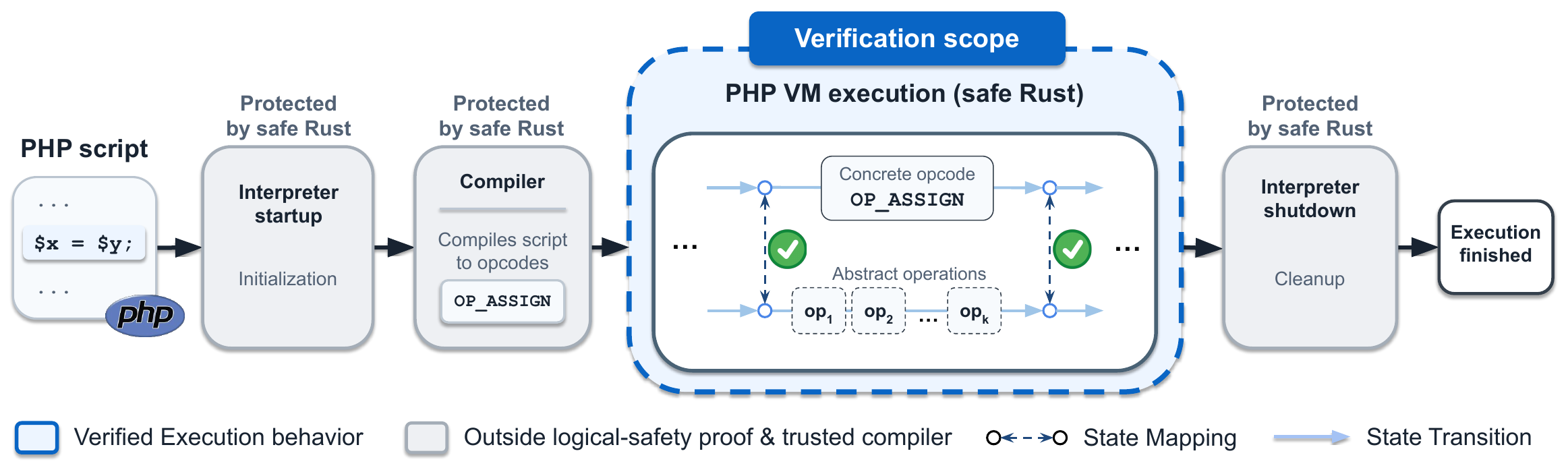}
    \caption{The overall protection scope of this paper, shown as different stages of a PHP execution.}
    \label{fig:scope}
\end{figure*}

\subsection{Logical Memory Safety}
\label{sec:logical-memory-safety-intuition}

As Section~\ref{sec:background} briefly illustrated, an interpreter can access an unintended program value without violating native memory safety. Such errors can undermine the logical validity of program values an operation is permitted to access. Preventing these errors requires an additional notion of safety at the level of the program's logical memory.

Unlike native memory safety, the security significance of a logical memory error depends on the specific application. In an image parser, reading an incorrect pixel within the bounds of an image may merely produce an incorrect result. But in a language interpreter, accessing the wrong value may expose sensitive information or corrupt program state. Thus, enforcing bounds on individual Rust container objects is not sufficient. The relevant safety property should reflect which values the program's operations are permitted to access, modify, and share its content legitimately.

Recent work has also recognized limitations of native memory-safety guarantees. CMASan~\cite{hong2025cmasan} addresses memory safety at the level of custom memory allocators, while a recent analysis of verification artifacts~\cite{mathur2026verified} highlights the importance of accessing live program objects and checking logical bounds inside native memory objects. These illustrate the need for stronger memory safety properties at finer abstraction levels than just native memory objects.

For language interpreters, the language semantics provides a basis for defining such properties precisely: they constrain which values each operation may read or modify and how it may modify sharing relationships. We therefore derive logical memory safety for PHP from PHP memory semantics, focusing on these two aspects of interpreter operations.

\subsection{Logical Memory Safety for PHP}
\label{sec:php-logical-memory-safety}

The execution of a PHP script can be represented as a sequence of
memory-state transitions,
$s_0 \xrightarrow{a_1} s_1 \xrightarrow{a_2} s_2
\xrightarrow{a_3} \cdots$.
Each state $s_i$ represents PHP variables, values, and their sharing
relationships, while each action $a_i$ represents the memory effects
of an interpreter operation, such as assignment, computation, or a
function call.
We define logical memory safety through two properties of these
transitions.

\noindent\textbf{(a) Frame property.}
An operation should (a) depend only on part of the state reachable from its
operands and (b) modify only the state it is permitted to change.
For example, the assignment \texttt{\$x = \$y + 1} should neither
depend on nor modify an unrelated variable \texttt{\$z}.
More precisely, for each transition $s \xrightarrow{a} s'$, we require:
\begin{itemize}
    \item \textbf{Read confinement:}
    Changing the state outside the region reachable from $a$'s
    operands must not affect the transition's result or its
    permitted effects.

    \item \textbf{Write confinement:}
    The transition must not modify state outside the region
    that $a$ is permitted to affect.
\end{itemize}

\noindent\textbf{(b) Sharing integrity.}
The frame property constrains memory accesses and modifications,
but does not constrain how an operation changes sharing relationships.
Consider an incorrect implementation of the assignment \texttt{\$x = \$y} that
creates an alias between the two variables instead of copying
the value.
Such an implementation will not violate the frame property because both
variables are operands of the operation.
However, a subsequent assignment to \texttt{\$y} would unexpectedly
change \texttt{\$x}.
Frame property will still not be violated at this second operation,
since the incorrect alias now makes \texttt{\$x} reachable
from \texttt{\$y}.

In other words, the frame property, may not, by itself, prevent unintended sharing.
This, in turn, may extend subsequent operations' access to values beyond
what PHP semantics permits.
We therefore require:
\begin{itemize}
    \item \textbf{No unintended sharing:}
    A transition must not introduce sharing relationships its operation does not permit.

    \item \textbf{No stale sharing:}
    A transition must remove sharing relationships
    that its operation requires it to remove.
\end{itemize}

Together, the frame property and sharing integrity define logical
memory safety for PHP.
Enforcing them requires a precise representation of the state
reachable from each operation's operands and the sharing
relationships that the operation may change.

\subsection{Abstract PHP Machine Model}
\label{sec:model-choice}

The preceding subsection outlines what logical memory safety should guarantee.
Making these guarantees precise requires identifying, for each operation, the state it may depend on or modify and the sharing relationships it may introduce or must remove.
Specifying these requirements consistently across PHP instructions is challenging.
Arithmetic operations read operands and write results, assignments may change sharing according to value types, and function calls additionally manage execution frames.

A full functional specification of PHP instructions would determine these memory behaviors, but specifying and verifying computation details would add obligations beyond our safety goal.
We instead define an abstract PHP machine that specifies the memory effects of a small set of core operations while leaving computation-specific functionality abstract.
The model also constrains behavior that safety alone leaves open, such as requiring scalar assignment to copy the source value.
Its rules provide a basis for establishing the frame property and sharing integrity without requiring full functional verification.

\noindent\textbf{Modeled state.}
To determine which values an operation may access or modify,
the model needs to distinguish PHP's different forms of sharing.
For example, \texttt{\$x =\& \$y} makes two variables aliases of
the same value, whereas \texttt{\$q = \$o}, when \texttt{\$o}
holds an object, makes two independent variables refer to the
same object.
These relationships have different effects: assigning a new
value to an alias affects both variables, while modifying a
shared object's properties is visible through both object variables.
Following the PHP specification, we therefore distinguish
variable slots, which identify variables or array elements/object
properties; value storage, which holds values and may be shared
by aliased slots; and heap storage, which represents arrays and
objects.
The connections between these storage locations determine
both operand reachability and sharing relationships.

\noindent\textbf{Details abstracted away.}
Our abstract machine specification models memory effects of operations
without stating their full functionality.
The exact results of arithmetic, string operations,
and other computations remain unspecified.
Their accesses to and modifications of the program state are restricted to states reachable from their operands.
The resulting abstract machine specifies the permitted memory-state changes of PHP operations, providing the basis for verifying the concrete Rust implementation.

\subsection{Verifying the Concrete Rust Implementation}
\label{sec:verification-overview}

The abstract PHP machine specifies the permitted memory-state
changes, but its safety alone does not guarantee the safety of
the concrete Rust interpreter.
We establish this connection in two steps.
First, we prove that every transition of the abstract machine
satisfies the frame property and sharing integrity.
Second, we prove that the verified execution branches of the concrete Rust interpreter refine
the abstract machine~\cite{abadi1991refinement}.

To establish refinement, we relate concrete states of our Rust implementation of the PHP interpreter to the states of the abstract PHP machine model and
show that the memory-state changes
performed by the operations in the concrete interpreter correspond to permitted
abstract transitions.
This allows the abstract machine's safety guarantees to
transfer to the concrete implementation our proof covered,
without requiring
a proof of full functional correctness.
Both proofs are formalized in Verus and pass the check~\cite{lattuada2023verus}.
\SubmissionOnly{
We provide these proofs in our artifact~\footnote{\href{https://drive.google.com/file/d/1AGalSVpl8LBApmZoTO4mZlhj1_L3gZmS/view?usp=sharing}{Anonymized artifact repository}}.
}
\ArxivOnly{
The formal proofs and accompanying artifacts will be made publicly available soon.
}

\noindent\textbf{Scope of verification.}
Our safety property mainly targets the execution stage of the translated interpreter.
Interpreter startup, PHP script compilation, and shutdown are outside
the modeled execution and are therefore not covered by the
logical memory-safety proof.
These stages are still protected by safe Rust, so native memory safety is established by construction.
We assume the PHP compiler correctly compiles the source scripts into correct opcode sequences, and do not formally verify its correctness.
Figure~\ref{fig:scope} summarizes the scope of this paper's verification effort.
In Section~\ref{sec:model}, we formally define the abstract machine and prove
its safety properties.
Section~\ref{sec:refinement-verification} presents the refinement verification of the
concrete Rust implementation.

%% file: sections/04-memory-model.tex
\section{The Formal Abstract PHP Machine Model}
\label{sec:model}

We now formalize the abstract PHP machine model used to state and enforce PHP memory-model safety,
based on the storage relationships specified by the PHP specification~\cite{php-memory-spec}.

\subsection{Abstract PHP machine}
\label{sec:abstract-machine}

\begin{definition}[Abstract PHP machine]
\label{def:abstract-machine}
An abstract PHP machine is a labeled transition system:
\begin{equation}
    \mathcal{M}=(S,OP,s_{\emptyset},\longrightarrow,\mathcal I),
    \label{eq:abstract-machine}
\end{equation}
$S$ is the set of well-defined PHP memory states.
$OP$ is the set of abstract operations.
$s_{\emptyset}$ is the empty initial state.
$\xrightarrow{a}\,\subseteq S\times S$ specifies the effect of operation $a\in OP$.
The interpretation $\mathcal I$ determines the results
of computations without fixing their functionality in the model. It is
fixed for each machine instance and is defined further in
Section~\ref{sec:transition}.
An execution is a sequence
$s_{0}\xrightarrow{a_1}s_1\xrightarrow{a_2}s_2\cdots$ such that $s_0 = s_{\emptyset}$ and for every $i$, $(s_i, s_{i+1}) \in \xrightarrow{a_i}$.
\end{definition}

\subsection{PHP Memory States}
\label{sec:memory-states}

The PHP machine distinguishes three kinds of storage, corresponding to the
PHP specification's separation of variables and values:
\emph{variable slots} (VSlots), \emph{value storage locations} (VStores),
and \emph{heap storage locations} (HStores).

A VSlot holds a reference to a VStore;
a VStore holds either a scalar or a handle to an HStore;
and an HStore represents an array or object containing VSlots indexed by keys.
These relationships permit two forms of sharing: multiple VSlots can
point to the same VStore, and distinct VStores can hold handles to the
same object HStore.

\input{figures/php_state_example}

For example, Figure~\ref{fig:state-example} shows the state after
\texttt{\$x = 1; \$y = \&\$x;} and
\texttt{\$o = new C; \$o->m = 3; \$q = \$o;}.
Variables \texttt{\$x} and \texttt{\$y} alias one VStore, whereas
\texttt{\$o} and \texttt{\$q} hold distinct VStores pointing to the
same object HStore.

Let $\mathsf{Var}$, $\mathsf{Key}$, $\mathsf{Store}$, and $\mathsf{Heap}$
be pairwise disjoint sets of variable names, keys, VStore identities, and HStore identities, respectively.
The location of a VSlot is identified either by a variable or by a key in an HStore:
\begin{equation}
    \mathsf{Slot}=\mathsf{Var}\uplus(\mathsf{Heap}\times\mathsf{Key}).
    \label{eq:slot-domain}
\end{equation}
We write $h.k$ for the slot $(h,k)$, and let $\mathsf{Scalar}$
denote the scalar values, including \texttt{null}.  These domains are
disjoint from one another.

\begin{definition}[State]
\label{def:state}
A state $s=(\mathrm{pt}_s,\mathrm{val}_s,\mathrm{kind}_s)$ comprises three
finite partial maps:
\begin{equation}
\begin{aligned}
 \mathrm{pt}_s &: \mathsf{Slot}\rightharpoonup\mathsf{Store},\\
 \mathrm{val}_s &: \mathsf{Store}\rightharpoonup
                         (\mathsf{Scalar}\uplus\mathsf{Heap}),\\
 \mathrm{kind}_s &: \mathsf{Heap}\rightharpoonup
                         \{\mathsf{arr},\mathsf{obj}\}.
\end{aligned}
\label{eq:state}
\end{equation}
The first map identifies the VStore referenced by each VSlot; the second
records the scalar or handle contained in each VStore; and the third
identifies each HStore as an array or an object.  An HStore $h$ contains
precisely those slots $h.k$ for which $\mathrm{pt}_s(h.k)$ is defined.
We write $\mathsf{Slot}_s$, $\mathsf{Store}_s$, and $\mathsf{Heap}_s$
for the respective domains of these maps.

A state is \emph{well-defined} when the three mappings are consistent,
i.e., every VSlot points to an existing VStore,
every HStore handle refers to an existing HStore,
every slot $h.k$ has an existing parent HStore,
and every array HStore has at most one VStore holder.
The following discussion is based on well-defined states, unless stated otherwise.
We take $S$ to be the set of well-defined states and $s_{\emptyset}=(\emptyset,\emptyset,\emptyset)$.
\end{definition}

To precisely state our logical memory safety definition,
we will need to identify the storage
{\em reachable} from a set of VSlots, through
VSlot pointers, HStore handles, and the slots contained in each HStore.
More formally, for $X\subseteq\mathsf{Slot}$, define $\mathrm{reach}_s(X)$
as the least set $R$ of storage identities such that $X\subseteq R$
and the following conditions hold:
\begin{equation}
\begin{aligned}
 \ell\in R\cap\mathsf{Slot}_s
    \implies &\mathrm{pt}_s(\ell)\in R,\\
 v\in R\cap\mathsf{Store}_s,\ \mathrm{val}_s(v)=h
    &\implies h\in R,\\
 h\in R\cap\mathsf{Heap}_s,\ h.k\in\mathsf{Slot}_s
    &\implies h.k\in R.
\end{aligned}
\label{eq:reach-edges}
\end{equation}
We write $s\upharpoonright_R$ for the \emph{restriction} of
the state $s$ on the identifiers in $R$,
i.e., the three maps restricted to $R$,
together with the number of holders of each
VStore and HStore in $R$, counted in the whole state.

\subsection{Operations and Transition Relation}
\label{sec:transition}

An operand $p$ denotes either a variable, an array element, or an object
field reached by a sequence of keys indexing into another object. The syntax of the operations is
\begin{equation}
\begin{aligned}
 p ::={}& x\mid p[k],\\
 a ::={}& \mathsf{create}\ p
       \mid p:=\mathsf{new}\ \kappa
       \mid p:=q
       \mid p=\&q\\
      &\mid p:=\mathsf{clone}\ q
       \mid \mathsf{unset}\ p\\
      &\mid p:=\mathsf{compute}(p_1,\ldots,p_n)\ \mathsf{by}\ o,
\end{aligned}
\label{eq:actions}
\end{equation}
where $x\in\mathsf{Var}$, $k\in\mathsf{Key}$,
$\kappa\in\{\mathsf{arr},\mathsf{obj}\}$, and $o$ identifies a scalar
computation.

The seven operations respectively create a slot,
allocate an empty array or object,
perform value assignment,
perform reference assignment,
clone an object,
remove a slot,
and compute a scalar.
Function calls and control flow are not separate operations; their
executed bodies are represented by the operations they perform.

To apply an operation, an existing operand must resolve to a VSlot.
Its denotation is defined recursively:
\begin{equation}
\begin{aligned}
 \llbracket x\rrbracket_s &= x,
     &&x\in\mathsf{Slot}_s,\\
 \llbracket p[k]\rrbracket_s &= h.k,
     &&\mathrm{val}_s(\mathrm{pt}_s(\llbracket p\rrbracket_s))=h,\\
 &&&h.k\in\mathsf{Slot}_s.
\end{aligned}
\label{eq:operand-denotation}
\end{equation}

\noindent\textbf{Representative transition rules.}
A transition $s\xrightarrow{a}s'$ applies action $a$ to state $s$.
We show the rules for value assignment, reference assignment, and
scalar computation; the full set of rules is given in
Appendix~\ref{app:model}.
Let us fix the following notation:
$\ell_p=\llbracket p\rrbracket_s$,
$\ell_q=\llbracket q\rrbracket_s$,
$v_p=\mathrm{pt}_s(\ell_p)$, and $v_q=\mathrm{pt}_s(\ell_q)$. \\

\noindent
\underline{\it Value assignment} replaces the content of the destination VStore
without changing the destination VSlot's pointer. It copies scalars,
duplicates arrays, and shares object handles:
\begin{equation}
 \frac{(c,\Delta)=\mathrm{copy}_s(\mathrm{val}_s(v_q))}
 {s\xrightarrow{p:=q}
       (s\uplus\Delta)[v_p\mapsto c]}
 \quad\textsc{Assign}
 \label{eq:assign-rule}
\end{equation}
Here, $s[u\mapsto w]$ updates the appropriate state map at storage identity $u$, and $s\uplus\Delta$ extends $s$ with fresh storage entries $\Delta$.
The copying operation $\mathrm{copy}_s(c)=(c',\Delta)$ returns the
copied content and any fresh storage allocated:
\begin{equation}
 \mathrm{copy}_s(c)=
 \begin{cases}
  (c,\emptyset), &c\in\mathsf{Scalar}
                \text{ or }\mathrm{kind}_s(c)=\mathsf{obj},\\
  \mathrm{dup}_s(c), &\mathrm{kind}_s(c)=\mathsf{arr}.
 \end{cases}
 \label{eq:copy}
\end{equation}
Scalars and object handles are copied unchanged, whereas
$\mathrm{dup}_s(h)=(h',\Delta)$ allocates a fresh array HStore and
recursively copies its non-reference elements, including nested arrays.
The precise duplication rules are given in Appendix~\ref{app:model}. \\

\noindent
\underline{\it Reference assignment} redirects the destination VSlot
to the source VStore, establishing an alias regardless of the kind of
value held there:
\begin{equation}
 \frac{
       }
 {s\xrightarrow{p=\&q}
       s[\ell_p\mapsto v_q]}
 \quad\textsc{Bind}
 \label{eq:bind-rule}
\end{equation}

\noindent
\underline{\it Computation operation}.
Let $\{\ell_i=\llbracket p_i\rrbracket_s\}_{i=1}^{n}$ be the
source slots for each of the operands. The computation receives only their reachable state:
\begin{equation}
\begin{aligned}
 &R=\mathrm{reach}_s(\{\ell_1,\ldots,\ell_n\}),\\
 &\mathrm{view}_s(\ell_1,\ldots,\ell_n)
   =((\ell_1,\ldots,\ell_n),s\upharpoonright_R).
\end{aligned}
\label{eq:compute-view}
\end{equation}
Intuitively, the view is everything the computation may look at: its
source operands and the storage they reach.
The interpretation $\mathcal I(o)$ takes this view as input and returns a scalar
$d$.
With $v_p=\mathrm{pt}_s(\llbracket p\rrbracket_s)$, the computation rule is:
\begin{equation}
 \frac{d=\mathcal I(o)(\mathrm{view}_s(\ell_1,\ldots,\ell_n))}
 {s\xrightarrow{p:=\mathsf{compute}(p_1,\ldots,p_n)\ \mathsf{by}\ o}
       s[v_p\mapsto d]}
 \quad\textsc{Compute}
 \label{eq:compute-rule}
\end{equation}
The model does not prescribe the result of a particular computation,
but the fixed function $\mathcal I(o)$ must depend only on the view of its
operands, not on the storage names:
$\mathcal I(o)(w)=\mathcal I(o)(\rho(w))$ for every view $w$ and every
one-to-one renaming $\rho$ of VStore and HStore names.
So a computation can tell whether two operands lead to the same storage,
but it cannot use a name as a value.

\subsection{Precise Logical Memory Safety for PHP}
\label{sec:logical-memory-safety}

We are now ready to define the frame property and sharing integrity from Section~\ref{sec:overview}
formally.
For an operation $a$, let $X_s(a)$ be the ordered sequence of VSlots
identified by its operands in state $s$. When $a$ is of the form
$\mathsf{create}\ p$, we take $X_s(a)$ to contain just the slot created by $a$.

\begin{definition}[Frame property]
\label{def:frame-property}
For a transition $s\xrightarrow{a}s'$, let
$N=\mathrm{reach}_s(X_s(a))$ and $N'=\mathrm{reach}_{s'}(X_s(a))$.
be its before and after \emph{name sets}. The transition
satisfies the \emph{frame property} if:
\begin{enumerate}
 \item \emph{Read confinement.}
       For every one-to-one renaming $\rho$ of VStore and HStore names,
       and every transition $t\xrightarrow{a}t'$ with
       $X_t(a)=\rho(X_s(a))$, let $M,M'$ be its corresponding name sets. Then
       \begin{equation}
        t\upharpoonright_M=\rho(s\upharpoonright_N)
        \Longrightarrow\quad
        t'\upharpoonright_{M\cup M'}
            \cong_\rho
        s'\upharpoonright_{N\cup N'}.
       \label{eq:frame-read}
       \end{equation}
       Here, $\cong_\rho$ permits a consistent renaming of VStore and HStore identities.
       Existing identities are renamed according to $\rho$, while fresh identities
       are matched between the two transitions. When $\rho$ is the identity,
       only fresh identities may be renamed.

       That is, changing unrelated input state leaves the operation's result and
       effects in its named region unchanged up to renaming.
 \item \emph{Write confinement.} If
       $R=(\mathrm{dom}(s)\cup\mathrm{dom}(s'))\setminus(N\cup N')$,
       where $\mathrm{dom}(s)$ is the union of the state's storage
       domains, then
       \begin{equation}
           s\upharpoonright_R=s'\upharpoonright_R.
           \label{eq:frame-write}
       \end{equation}
       That is, the operation should not modify storage outside the named region.
\end{enumerate}
\end{definition}

Let us now formalize sharing integrity using the notion of connections. A \emph{connection} links a VSlot to the VStore it points to, or a VStore
to the HStore whose handle it holds. Let $\mathrm{Conn}(s)$ be the set of
connections in state $s$:
\begin{equation}
\begin{aligned}
 \mathrm{Conn}(s)
  =&\{(\ell,\mathrm{pt}_s(\ell))\mid \ell\in\mathsf{Slot}_s\}\\
  \cup&\{(v,h){\mid}v{\in}\mathsf{Store}_s, \mathrm{val}_s(v){=}h\in\mathsf{Heap}\}.
\end{aligned}
\label{eq:connections}
\end{equation}
Two VSlots are aliases when their links end at the same VStore, and two
VStores share an object when their links end at the same HStore.
For each operation $a$, let $\mathrm{Add}_s(a)$ be the connections it may
create and $\mathrm{Rm}_s(a)$ the connections it must remove
(as specified in Table~\ref{tab:allowed-connections}).

\begin{definition}[Sharing integrity]
\label{def:sharing-integrity}
A transition $s\xrightarrow{a}s'$ satisfies \emph{sharing integrity}
if
\begin{equation}
 \mathrm{Conn}(s')\subseteq
       (\mathrm{Conn}(s)\setminus\mathrm{Rm}_s(a))
          \cup\mathrm{Add}_s(a).
 \label{eq:sharing-integrity}
\end{equation}
Thus, each connection in $s'$ must either be an existing connection not
marked for removal or a connection that $a$ may create.
\end{definition}

In the table, $\mathrm{Conn}_s(u)$ denotes the connections that start at storage $u$.
We use
$\ell=\llbracket p\rrbracket_s$,
$v=\mathrm{pt}_s(\ell)$, and
$v'=\mathrm{pt}_s(\llbracket q\rrbracket_s)$;
for $\mathsf{create}\ p$, $\ell$ is the slot it creates.
An operation that replaces the content of a VStore must cut the link
from that VStore to its old HStore, while the links of its VSlots stay.
In contrast, reference assignment redirects a VSlot: it must cut the
VSlot's old link, and it may add a link to the source VStore.
Value assignment and cloning may add a link from the destination VStore
to the copied value, which can be a shared object handle as in
Figure~\ref{fig:state-example}, and the links inside the new copy.
These permitted links are denoted by $\mathrm{CopyConn}_s(a)$, and their
precise definition is in Appendix~\ref{app:model}.

\begin{table}[t]
\centering
\caption{The connections each operation may create
($\mathrm{Add}$) and must cut ($\mathrm{Rm}$).}
\label{tab:allowed-connections}
\begin{tabular*}{\columnwidth}{@{\extracolsep{\fill}}lll@{}}
\toprule
Operation $a$ & $\mathrm{Add}_s(a)$ & $\mathrm{Rm}_s(a)$\\
\midrule
$\mathsf{create}\ p$ & $\{(\ell,w)\mid w\notin\mathsf{Store}_s\}$ & $\emptyset$\\
$p:=\mathsf{new}\ \kappa$ & $\{(v,h)\mid h\notin\mathsf{Heap}_s\}$ & $\mathrm{Conn}_s(v)$\\
$p:=q$ & $\mathrm{CopyConn}_s(a)$ & $\mathrm{Conn}_s(v)$\\
$p=\&q$ & $\{(\ell,v')\}$ & $\mathrm{Conn}_s(\ell)$\\
$p:=\mathsf{clone}\ q$ & $\mathrm{CopyConn}_s(a)$ & $\mathrm{Conn}_s(v)$\\
$\mathsf{unset}\ p$ & $\emptyset$ & $\mathrm{Conn}_s(\ell)$\\
$\mathsf{compute}(\ldots)$ & $\emptyset$ & $\mathrm{Conn}_s(v)$\\
\bottomrule
\end{tabular*}
\end{table}

\begin{definition}[Logical memory-model safety for PHP]
\label{def:logical-memory-safety}
An execution of the abstract PHP machine $\mathcal M$ (Definition~\ref{def:abstract-machine}) is \emph{PHP memory-model safe} if every
transition satisfies the frame property (Definition~\ref{def:frame-property}) and sharing
integrity (Definition~\ref{def:sharing-integrity}).
\end{definition}

\begin{theorem}[Abstract-machine safety]
\label{thm:abstract-machine-safety}
For every fixed interpretation $\mathcal I$ that takes only the
specified operand view as input and does not depend on its names,
every execution admitted by
$\mathcal M$ is PHP memory-model safe.
\end{theorem}

The abstract machine and its safety properties are formalized in Verus,
together with a machine-checked proof of
Theorem~\ref{thm:abstract-machine-safety}; a proof sketch is provided in
Appendix~\ref{app:safety-proof}.

%% file: figures/php_state_example.tex
\begin{figure}[t]
\centering
\begin{tikzpicture}[
  >=Stealth,
  font=\scriptsize,
  link/.style={->, semithick},
  slot/.style={
    draw, rounded corners=1pt, fill=black!4,
    minimum width=8mm, minimum height=3.8mm, inner sep=1pt
  },
  store/.style={
    draw, rounded corners=1pt, fill=black!10,
    minimum width=10mm, minimum height=4.2mm, inner sep=1pt
  },
  field/.style={
    draw, rounded corners=1pt, fill=white,
    minimum width=10mm, minimum height=3.8mm, inner sep=1pt
  }
]

  \node[font=\bfseries\scriptsize] at (0, 1.55) {VSlots};
  \node[font=\bfseries\scriptsize] at (1.65, 1.55) {VStores};
  \node[font=\bfseries\scriptsize] at (3.90, 1.55) {HStore};

  \node[slot]  (x)  at (0, 1.00) {\texttt{\$x}};
  \node[slot]  (y)  at (0, 0.45) {\texttt{\$y}};
  \node[store] (vx) at (1.65, 0.73) {$v_1\!:\!1$};

  \draw[link] (x.east) -- (vx.west);
  \draw[link] (y.east) -- (vx.west);

  \node[slot]  (o)  at (0, -0.35) {\texttt{\$o}};
  \node[slot]  (q)  at (0, -0.95) {\texttt{\$q}};
  \node[store] (vo) at (1.65, -0.35) {$v_2$};
  \node[store] (vq) at (1.65, -0.95) {$v_3$};
  \node[store] (vm) at (1.65, -1.65) {$v_4\!:\!3$};

  \draw[link] (o.east) -- (vo.west);
  \draw[link] (q.east) -- (vq.west);

  \draw[fill=black!3, rounded corners=1.5pt]
    (3.15, -0.05) rectangle (4.65, -1.95);

  \node[font=\scriptsize] at (3.90, -0.38)
    {$h$: object \texttt{C}};

  \draw[densely dotted] (3.28, -0.62) -- (4.52, -0.62);

  \node[field] (prop) at (3.90, -1.55) {$h.\texttt{m}$};

  \draw[link] (vo.east) -- (3.15, -0.35);
  \draw[link] (vq.east) -- (3.15, -0.95);
  \draw[link] (prop.west) -- (vm.east);

\end{tikzpicture}
\caption{An abstract PHP memory state. Variables \texttt{\$x} and
\texttt{\$y} alias the same VStore $v_1$, whereas \texttt{\$o} and
\texttt{\$q} refer through distinct VStores to the same object HStore
$h$. The object property \texttt{m} is itself a VSlot pointing to $v_4$.}
\label{fig:state-example}
\end{figure}
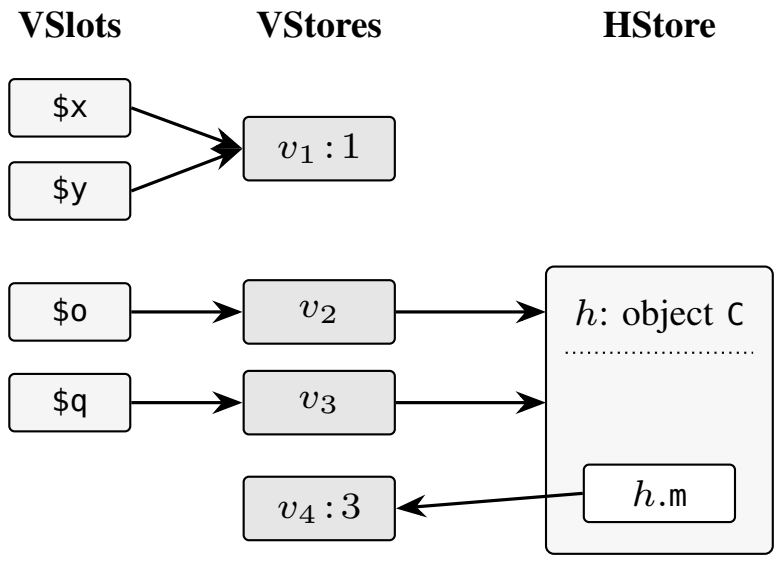

%% file: sections/05-refinement-verification.tex
\section{Refinement-Based Verification}
\label{sec:refinement-verification}

Section~\ref{sec:model} establishes that the abstract PHP machine satisfies
logical memory safety.
To transfer this guarantee to the Rust interpreter, we prove
that a restriction to a core set of its concrete memory operations refines the abstract machine~\cite{abadi1991refinement}.

The proof has three components.
First, we map concrete Rust states to abstract PHP states while
preserving PHP-visible observations.
Second, we identify invariants that ensure the concrete memory
representation remains well-defined.
Third, for each verified opcode branch, we prove that its Rust
implementation preserves these invariants and performs a
predetermined sequence of abstract operations.
We encode and check these obligations in Verus.

\subsection{Mapping Concrete States to Abstract States}
\label{sec:state-mapping}

We relate concrete Rust states to abstract PHP states using a mapping
$\alpha$ that preserves PHP values and sharing relationships.
Additionally, we define a representation invariant $\mathit{Inv}(c)$
that requires the concrete state $c$ to have a well-defined memory
representation.

\noindent\textbf{State mapping.}
The mapping must cover the interpreter's runtime state as well as PHP
variables and values. Interpreter operations access or modify program
counters, execution frames, and temporaries, so omitting these would
leave some accesses unverified. We map PHP variables and values
according to the storage model in Section~\ref{sec:model} and represent
runtime state using hidden abstract storage.

For each execution frame $d$, we map its program counter to a hidden
abstract variable $pc_d$, its variable table to an abstract array
$vars_d$, and its temporaries to hidden variables
$tmp_d[0], tmp_d[1], \ldots$. A PHP variable \texttt{\$x} in frame $d$
then corresponds to $vars_d[x]$. We similarly represent the argument
stack, pending calls, and call stack by hidden arrays, and the output
by a hidden string. These hidden storage locations cannot be named by
PHP programs, including through variable variables. Function and class
definitions and executable code are tracked separately as part of the
state correspondence.

A precise description of this mapping requires care. In particular,
each PHP operand must resolve to the same logical value in the concrete
and abstract states, with the same reference aliases and shared object
identities, up to consistent renaming of storage identities.
This correspondence prevents the mapping from justifying an unintended
access by interpreting a concrete handle as an unrelated abstract value.
We encode this correspondence in Verus.

\noindent\textbf{Representation invariant.}
The mapping relies on a well-defined concrete memory representation.
References and object handles must refer to live allocations, and their
reference counts must agree with the numbers of holders. An incorrect
release or copy may violate these conditions and invalidate the
correspondence.

The invariant $\mathit{Inv}(c)$ requires that (a) storage structures and
runtime tables are consistent, (b) all handles held by the interpreter
refer to live values, (c) reference and object counts equal their actual
numbers of holders, and (d) allocation identities and metadata are
consistent. We prove that (a) this invariant holds in the initial
concrete state and that (b) each verified opcode branch preserves it
at execution boundaries.

\subsection{Verifying Opcode Refinement}
\label{sec:opcode-refinement}

The state mapping and invariants already establish a basic level of safety, ruling out violations like dangling PHP values and refcounting errors.
We additionally constrain how
the concrete interpreter changes the memory state.

To establish this constraint, we prove that each verified
opcode performs only the memory-state changes permitted
by the abstract machine.

\noindent\textbf{Opcode specifications.}
We specify the expected memory effects of each opcode
as a predetermined sequence of abstract operations,
which we call its \emph{table entry}.
An opcode may have multiple entries for different execution
branches, such as different operand types.
Each entry is determined by the opcode and its execution
conditions, rather than chosen retrospectively to match
the concrete result.
We construct these entries manually using the abstract
operations defined in Section~\ref{sec:model}.

For example, consider a value assignment
\texttt{\$x = \$y} executed by frame $d$ at instruction $pc_d$.
Its table entry is
\[
    [\mathsf{create}\ \texttt{\$x}];\quad
    \texttt{\$x} := \texttt{\$y};\quad
    pc_d \leftarrow pc_d + 1.
\]
Here, \texttt{\$x} and \texttt{\$y} denote the corresponding
elements of the frame's abstract variable table,
and the brackets indicate that the slot is created
only if it does not already exist.
The entry creates the destination when necessary, copies
the source value according to the abstract assignment rule,
and advances the program counter ($pc_d \leftarrow pc_d + 1$ is a simplified format for the compute transition).
Other opcodes are specified similarly by sequences of
these abstract operations.

\noindent\textbf{Refinement proof.}
For each verified opcode branch, we prove in Verus that
its Rust implementation preserves the concrete representation
invariant and realizes its predetermined table entry.
Let $c$ and $c'$ be the concrete states before and after
executing a branch of opcode $\sigma$, and let
$\alpha(c)=s$ be the corresponding abstract state.
The proof establishes
\[
\begin{aligned}
    &\mathit{Inv}(c)
    \land \alpha(c)=s
    \land c \xrightarrow{\sigma} c'
    \\
    &\quad\Longrightarrow
    \mathit{Inv}(c')
    \land
    s \xrightarrow{a_1} \cdots
      \xrightarrow{a_k} \alpha(c'),
\end{aligned}
\]
where $a_1,\ldots,a_k$ are the operations prescribed
by that branch's table entry.
Because the entry is fixed independently of the concrete
post-state, an unintended memory modification cannot be
justified by selecting different abstract operations.
The state mapping also preserves PHP-visible values and
sharing relationships, preventing the proof from reinterpreting
an unrelated concrete value as a permitted abstract operand.

For computation operations, the abstract machine leaves
the exact result to the fixed interpretation $I$.
The correspondence with concrete computation functions
is established through verified or trusted contracts,
while the interpretation restricts dependencies to the
declared operands.
Thus, refinement constrains the memory dependencies
and effects of computations without requiring their
complete functional correctness.

\noindent\textbf{Composition.}
We prove the representation invariant for the initial
execution state.
Since every verified opcode branch preserves this invariant,
its post-state satisfies the precondition needed by the next
verified branch.
The individual refinement contracts can therefore be
composed across executions consisting of verified branches.
Together with the abstract machine's safety property,
this establishes logical memory safety for these executions.

\begin{figure}[t]
\centering
\resizebox{0.98\columnwidth}{!}{

\begin{tikzpicture}[
    >=Latex,
    font=\small,
    state/.style={
        draw=blue!50!black,
        rounded corners=2pt,
        minimum width=10mm,
        minimum height=6mm,
        inner sep=1.5pt,
        fill=gray!5
    },
    flow/.style={->, thick, black},
    callarc/.style={->, thick, blue!70!black},
    dashedcall/.style={dashed, thick, black},
    every node/.style={align=center}
]

\node[state] (s2) at (0,0) {$S_2$};
\node[state] (s4p) at (5.4,0) {$S_4'$};

\node[state] (s2p) at (0.8,1.5) {$S_2'$};
\node[state] (s3)  at (2.7,1.5) {$S_3$};
\node[state] (s4)  at (4.6,1.5) {$S_4$};

\draw[flow] (-1.7,0) -- node[above] {$op_i$} (s2.west);
\draw[flow] (s4p.east) -- node[above] {$op_{i+1}$} +(1.7,0);

\draw[flow] (s2p.east) -- node[above] {$op_j$} (s3.west);
\draw[flow] (s3.east)  -- node[above] {$op_{j+1}$} (s4.west);

\usetikzlibrary{calc}
\coordinate (cL) at ($(s2.east)+(0.55,0)$);
\coordinate (cR) at ($(s4p.west)+(-0.55,0)$);
\draw[dashedcall] (s2.east) -- (cL);
\draw[dashedcall] (cL) -- node[above=2pt] {\textsc{DoCall}} (cR);
\draw[flow] (cR) -- (s4p.west);
\filldraw[white] (cL) circle (1.5pt);
\filldraw[white] (cR) circle (1.5pt);

\draw[callarc] (cL) .. controls +(0,0.75) and +(-0.8,-0.8) .. (s2p.west);
\draw[callarc] (s4.east) .. controls +(0.9,-0.8) and +(0,0.75) .. (cR);

\node[state] (bs2) at (0,-2.1) {$S_2$};
\node[state] (bs4p) at (5.4,-2.1) {$S_4'$};

\node[state] (bs2p) at (0.8,-0.7) {$S_2'$};
\node[state] (bs3)  at (2.7,-0.7) {$S_3$};
\node[state] (bs4)  at (4.6,-0.7) {$S_4$};

\draw[flow] (-1.7,-2.1) -- node[above] {$op_i$} (bs2.west);
\draw[flow] (bs4p.east) -- node[above] {$op_{i+1}$} +(1.7,0);

\draw[flow] (bs2p.east) -- node[above] {$op_j$} (bs3.west);
\draw[flow] (bs3.east)  -- node[above] {$op_{j+1}$} (bs4.west);

\draw[callarc] (bs2.east) .. controls +(1.0,0.3) and +(-0.9,-0.8) ..
    node[left=3pt,pos=0.45] {\textsc{Call}} (bs2p.west);

\draw[callarc] (bs4.east) .. controls +(0.9,-0.8) and +(-1.0,0.3) ..
    node[right=3pt,pos=0.52] {\textsc{Return}} (bs4p.west);

\end{tikzpicture}
}

\caption{Modeling the function call operation by splitting it into call and return halves.}
\label{fig:split-call}
\end{figure}
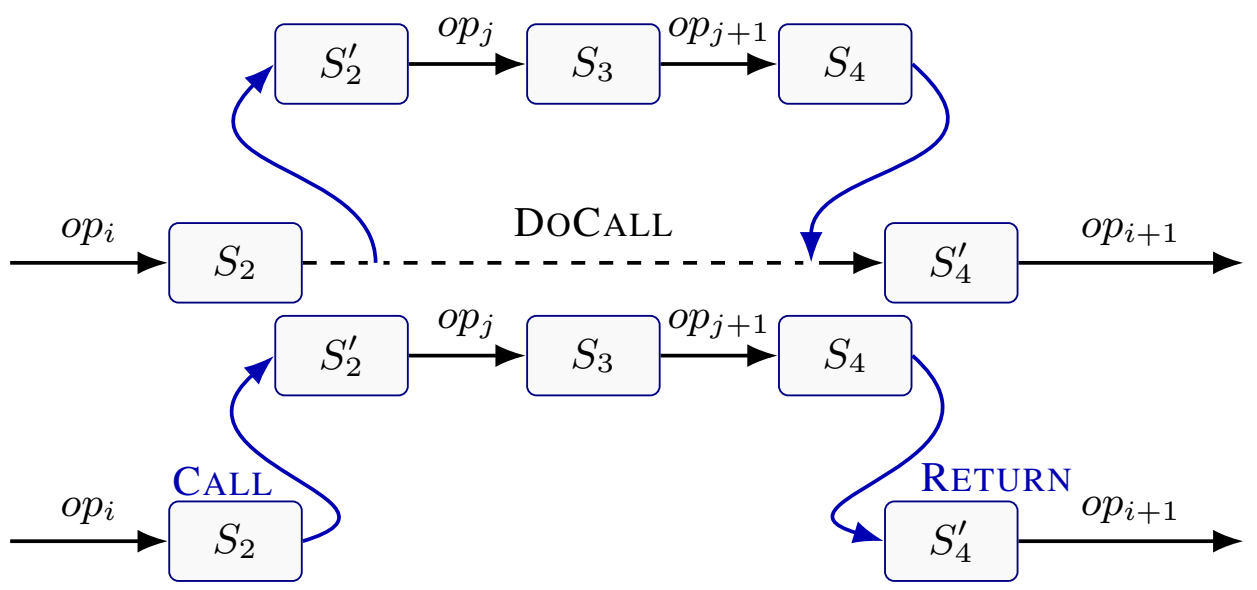

\subsection{Modeling and Verifying Function Calls}
\label{sec:verifying-function-calls}

Function calls modify execution frames, argument bindings,
and control flow.
We leave the function call effect outside the abstract PHP machine's
operation, and express these effects using its existing
memory operations.

\noindent\textbf{User-defined function calls.}
In the translated interpreter, the \texttt{DoCall} opcode
invokes a user-defined function in PHP by entering a new execution
frame.
The interpreter executes the function body in that frame,
then returns the result to the caller and releases the
callee's state.
Modeling the entire call as one transition would require
its specification to capture the combined effects of all
instructions in the function body.
However, the intervening instructions can be verified independently using the same opcode refinement method.
We therefore split the call into two steps:
the \emph{call half}, which ends at the callee's first
instruction, and the \emph{return half}, which ends at
the caller's next instruction, as shown in Figure~\ref{fig:split-call}.
The intervening execution is modeled by the ordinary
opcode transitions of the callee.

For example, when frame $d$ calls a function $f$ at
instruction $i$, the two halves have the following
schematic entries:

\[
\begin{aligned}
\textsc{Call}: \quad&
    \mathsf{create}\ vars_{d+1};\\
&   vars_{d+1} := \mathsf{new}\ \mathsf{arr};\\
&   \cdots\\
&   \mathsf{create}\ pc_{d+1};\\
&   pc_{d+1} \leftarrow 0.
\\[4pt]
\textsc{Return}: \quad&
    [\mathsf{create}\ tmp_d[r]];\\
&   tmp_d[r] := v;\\
&   \cdots\\
&   \mathsf{unset}\ vars_{d+1};\
        \mathsf{unset}\ pc_{d+1};\\
&   pc_d \leftarrow pc_d + 1.
\end{aligned}
\]

Here, $vars_{d+1}$ and $pc_{d+1}$ represent the
callee's variable table and program counter,
$v$ is the returned value, and $tmp_d[r]$ is the
destination temporary in the caller.
The operations that manage arguments, call-stack
entries, etc., are omitted.
The call half creates the callee's state and initializes
its program counter.
The return half transfers the result, releases the
callee's storage, and advances the caller's program counter.
The predetermined entries constrain the memory effects
of both halves, while the callee's instructions are
verified independently.

\noindent\textbf{Library function calls.}
Library functions, including built-ins and extensions,
are implemented in a host language such as C or Rust,
rather than in PHP.
They use the same argument-passing and return mechanisms
as user-defined functions, but their handlers execute
directly within \texttt{DoCall}, without dispatching
a sequence of PHP opcodes in a new execution frame.
Consequently, their memory effects are specified the same as ordinary opcode handlers.
For example, \texttt{strlen(\$s)} is specified to compute a result
from its argument, whereas \texttt{array\_push(\$a, \$v)} is permitted to
append a value to the array.

\subsection{Verification Scope}
\label{sec:verification-scope}

We used AI agents for writing Verus proofs.
The proof obligations are fully stated, but
due to token cost constraints, we have not verified all PHP instructions and libraries.
We finished the proof on the main execution branches of
approximately 60 of the translated interpreter's 80 opcode types,
covering core operations on scalar, array, reference, object, and function-call.
For each verified branch, we proved both preservation of the representation invariant and refinement against its predetermined abstract operations.

The proof depends on assumptions about the behavior of the surrounding environment, such as the library functions the interpreter calls into.
To demonstrate the feasibility,
we additionally prove invariant preservation for about 40 PHP library function calls.
We did not see obvious challenges to apply our proof techniques to cover other opcodes, language features, and library functions.
The detailed contracts, proof obligations,
trusted assumptions, and unsupported execution cases are reported in our artifact.

The full translation includes numerous extensions and a diverse set of opcodes, totaling over 60K lines of Rust code that need verification.
We focus our proof effort on the commonly used operations and extensions, which cover 60\% of the code lines in the core engine and 40\% in the extensions.
Finishing all the opcodes and extensions will likely cost an additional \$200K USD.

%% file: sections/06-implementation.tex
\section{Implementation}
\label{sec:impl}

\noindent\textbf{Target PHP version.}
We target PHP~5.0.0, considering its relatively small size,
while being similar enough to modern PHP in terms of core PHP value semantics.
The PHP 5.0 we target consists of 37K lines of code for the core interpreter engine, plus surrounding environment and extensions.

\noindent\textbf{Scope of the implementation.}
We fully reimplement the core interpreter engine in PHP~5.0.0~\cite{php2004source}.
Besides the core engine,
we also reimplement the necessary surrounding runtime and built-ins, including the language parser, the CLI and CGI server interfaces, and the selected standard libraries and extensions that are needed for running applications (see Section~\ref{sec:eval-func}).
In total, our translation roughly covers 115K lines of C code in the C PHP~5.0.0, and produces 77K lines of code in safe Rust.
The full list of scope in our reimplementation is shown in Appendix~\ref{app:impl}.

\noindent\textbf{Implementation approach.}
We translate the engine semi-automatically, with humans working in a loop with AI coding agents, following the abstract PHP memory machine in Section~\ref{sec:model}.
Our final Rust implementation contains no \code{unsafe} code.
The translation uses 6 direct dependencies: \code{slotmap}, \code{regex}, \code{regex-automata}, \code{fancy-regex}, \code{memchr}, and \code{miniz\_oxide}.
These dependencies support storage management, regular expressions, and compression.
Appendix~\ref{app:impl} provides further implementation details and the dependency breakdown.
We implement and evaluate the interpreters on an Intel Core i7-12700 with 64\,GB of RAM, running Ubuntu 24.04.3.

%% file: sections/07-Evaluation.tex
\section{Evaluation}
\label{sec:eval}

We answer $4$ main empirical research questions:

\begin{enumerate}
    \item How many C PHP memory vulnerabilities could potentially persist in a safe Rust translation?
    \item Does the PHP engine rewritten in Rust behave safely against inputs previously triggering memory bugs in C?
    \item Does the Rust engine preserve the functionality of the original C engine?
    \item What is the impact on performance of rewriting from C into safe Rust?
\end{enumerate}

%% file: sections/07-findings-security.tex
\subsection{Evaluation I: Safety}
\label{sec:eval-security}

Logical memory errors exist within the defined behavior of the C language.
Even if Rust eliminates all undefined behaviors in C, these can persist.
Therefore, we want to understand how frequently the PHP engine in C violates logical memory safety.

We conduct a conservative measurement of how many logical bugs could survive in safe Rust.
We focus on the violations of 3 PHP machine rules:
(a) the state is well-defined, where we check the liveness of the PHP values and the correctness of the reference counting.
(b) the frame property violation, where we check if the operation writes into values without permission.
(c) the sharing integrity, where we detect both unexpected binding and unexpected unbinding.
Among the collected 173 memory vulnerabilities (Section~\ref{sec:bug-landscape}), we find 56 violations in total.
Table~\ref{tab:violation} shows the breakdown per rule.

These PHP-level memory-state inconsistencies arise before an invalid native memory access.
For instance, holding a dangling pointer without dereferencing it does not violate C memory safety.
However, holding a dangling PHP value across operations is dangerous in the PHP interpreter.
If the next operation doesn't check the liveness of the value and accesses it, it triggers a use-after-free.
A careless Rust translation may preserve the behavior as holding a dangling index into the value container, as our motivating example demonstrates (Section~\ref{sec:motivation}).

\begin{table}[t]
\centering
\footnotesize
\setlength{\tabcolsep}{3pt}
\caption{Violations of the abstract machine by the C engine on the 173 vulnerability trigger inputs, per rule.}
\label{tab:violation}
\begin{tabular*}{\columnwidth}{@{\extracolsep{\fill}}l r r r@{}}
\toprule
Rule & \makecell{Memory\\(142)} & \makecell{Logic\\(31)} & Total \\
\midrule
State is well-defined      & 32 & 0  & 32 \\
No unlicensed sharing    & 6  & 14 & 20 \\
No unlicensed unsharing  & 0  & 4  & 4  \\
Writes within the frame  & 2  & 4  & 6  \\
\midrule
Union                 & 38 & 18 & 56 \\
\bottomrule
\end{tabular*}
\end{table}

{\bf Result I:} We aim to achieve both native memory safety and logical memory safety for the translated PHP interpreter.
We test whether the translation indeed behaves safely on all the 173 memory-related vulnerabilities.

After analyzing the memory access behavior of our Rust PHP engine,
we find \emph{no} native memory violations or logic memory violations on these vulnerability-triggering inputs.
The Rust translation behaves gracefully on all of these vulnerability-triggering inputs with no panic.
158 of them execute normally; the remaining 15 are stopped by the translated interpreter with a PHP error message.

\begin{tcolorbox}[
  colback=blue!4,
  colframe=blue!35!black,
  boxrule=0.5pt,
  arc=1pt,
  left=6pt,
  right=6pt,
  top=5pt,
  bottom=5pt
]

Logical memory violations happen frequently in our collected memory vulnerabilities.
Our translation behaves safely on all these vulnerabilities, without native memory violations or logical memory violations.
\end{tcolorbox}

%% file: sections/08-findings-performance.tex
\subsection{Evaluation II: Performance}
\label{sec:eval-perf}

Prior approaches to full memory safety, such as direct instrumentation of C code, incur large overheads of over $50\%$~\cite{simpson2013memsafe}.
We ask whether the approach of rewriting to safe Rust offers a compelling alternative.

\noindent\textbf{Benchmark and metrics.} We measure the performance mainly based on the standard PHP performance benchmark, \texttt{bench.php}, independently provided in the C PHP repository~\cite{php2004bench}.
It contains 18 separate workloads,
each of which runs in less than $1$ second on the C engine in our environment.
To measure long-running scenarios where PHP memory reclamation starts to work,
we extend each workload to run for roughly one minute by expanding the loop number without touching the code logic.
We call this extended version \texttt{bench-long.php}.
We also leverage our functionality tests (Section~\ref{sec:eval-func}) for additional performance evidence.

We use \emph{wall time}, i.e., the end-to-end time cost of execution on one script, as our main performance measurement.
We repeat the evaluation 3 times, and take the minimum run for both C and Rust.
We record the total wall time for each engine on each benchmark; its ratio is the runtime overhead.
Both C and Rust are compiled with the optimization flag described in Appendix~\ref{app:impl}.

\begin{figure}[t]
  \centering
  \includegraphics[width=\columnwidth]{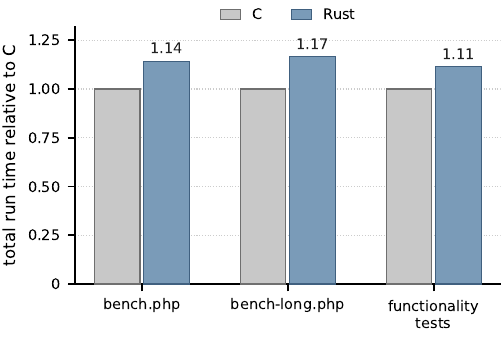}
  \caption{Total runtime comparison.}
  \label{fig:overhead}
\end{figure}

{\bf Result II:}
Figure~\ref{fig:overhead} shows the runtime comparison on 3 benchmarks.
On \texttt{bench.php}, the Rust runs at overall 1.14$\times$ C in wall time.
This ratio remains stable in the long-running scenario: the overhead on \texttt{bench-long.php} is around 17\%.
On the functionality tests, the overhead is roughly 11\%.

To better understand whether the performance overhead is inherent to Rust or stems from our design and implementation choices,
we conduct multiple fine-grained performance analyses.

First, we measure how much each stage contributes to the run time.
From the start of the interpreter to the end of execution, a run goes through four stages: startup, script parsing and compilation, execution, and shutdown.
We instrument both engines to time each stage of a run.
Both engines cost about the same at startup, and the Rust engine costs 6$\times$ more for compilation.
The compilation slowdown comes from a constant factor of about 0.2~ms per script, which dominates for small scripts.
As the script size grows, the compilation time overhead approaches about 2$\times$.
Overall, the execution stage takes more than 95\% of the total runtime on \texttt{bench.php} in both engines,
while the PHP compilation takes less than 1\% of the total wall time.
The runtime slowdown mainly comes from the execution stage.

To find exactly where execution slows down or speeds up,
we write 21 micro-benchmarks targeting 6 core PHP language features, shown in Table~\ref{tab:micro}.
We measure wall time as for \texttt{bench.php}, and count instructions with callgrind~\cite{valgrind_callgrind}.
We provide an empirical explanation for each group.

\begin{table}[t]
\centering
\footnotesize
\setlength{\tabcolsep}{3pt}
\caption{The 21 micro-benchmarks: the Rust engine's wall time and instruction count relative to C.}

\label{tab:micro}
\begin{tabular*}{\columnwidth}{@{\extracolsep{\fill}}l r r@{}}
\toprule
Group & Wall time & Instructions \\
\midrule
Operation Dispatch & 0.90$\times$ & 1.03$\times$ \\
Arithmetic      & 0.93$\times$ & 1.02$\times$ \\
Strings         & 1.18$\times$ & 1.22$\times$ \\
References      & 1.19$\times$ & 1.14$\times$ \\
Objects         & 1.29$\times$ & 1.25$\times$ \\
Function Calls  & 1.25$\times$ & 1.18$\times$ \\
\bottomrule
\end{tabular*}
\end{table}

\emph{Main dispatch loop and arithmetic} run faster in the Rust engine.
The speedup comes mainly from a design choice in how an operation reads its operands.
The C PHP engine runs a separate fetch operation for every variable it reads,
whereas our translation looks the variable up inside the operation that uses it.
In this way, the translation saves about three-quarters of the instructions for reading the operands.

\emph{String} operations cost 18\% run-time overhead in the Rust translation.
The slowdown comes from the memory management design.
In the C engine, a string allocation first tries to reuse a freed block from a small cache before it reaches the system allocator,
while the Rust engine allocates directly through the system allocator.
As a result, in the benchmark with frequent string concatenation, the Rust engine calls the system allocator 5 times more often than the C engine.
There is no fundamental reason that the safe Rust translation cannot express the same optimization trick as in C.
However, such designs can complicate the verification.
We leave this optimization as future work.

\emph{References and objects} carry the main slowdown.
We find two main causes, and both are design choices for safety in our translation.
First, the C PHP engine distinguishes a reference from an ordinary value by an \texttt{is\_ref} mark, and creating a reference just flips this mark.
This design is prone to reference-bookkeeping errors.
Our translation stores a reference as a separate type.
Creating or removing a reference therefore costs more instructions in converting value between the ordinary type and the reference type.
In the benchmark that creates and removes references, this accounts for about half of the slowdown.

Second, while an operation runs, it may re-enter PHP code, such as an object's destructor, and that code may change or free the variables the operation is using.
Our translation therefore copies the object before the operation starts, and frees objects that are no longer used only after the operation ends.
This keeps the object alive during the operation and avoids the conflict.
This extra copy costs about 11--13\% more instructions than the C engine on a method call.
Both designs aim to make the verification easier: Rust data types deliver the guarantees the verification relies on.

\emph{Calls} slow down in Rust mainly because of an implementation choice.
A function call switches the current call frame, and the C engine does so by exchanging pointers to the frame's data.
To keep the verification simpler, our Rust translation keeps the same design, and switches frames on every call and return by transferring the ownership of the frame.
Such ownership transfer costs about 50\% more instructions than a raw pointer copy in C.
A better design in Rust would keep all frames in one container and switch by index, without transferring ownership.
Similar to the customized cache mentioned in string operations, this extra layer of frame container would make verification more complex.
We leave it as future work.

Finally, we measure the overhead of the run-time checks of the Rust language.
These include the bounds checks on container accesses and the liveness (generation) checks on arena accesses.
On \texttt{bench.php}, turning off these run-time checks on the hot path gives a speedup of about 1\% overall.

\begin{tcolorbox}[
  colback=blue!4,
  colframe=blue!35!black,
  boxrule=0.5pt,
  arc=1pt,
  left=6pt,
  right=6pt,
  top=5pt,
  bottom=5pt
]

Our protection comes with a modest runtime overhead of 15\% on the PHP performance benchmark.
The overhead seems mainly from design and implementation choices, with a small part due to Rust's runtime checks.
\end{tcolorbox}

%% file: sections/09-findings-functionality.tex
\subsection{Evaluation III: Functionality}
\label{sec:eval-func}

The safe Rust translation should still match the original C engine in its PHP interpretation functionality.

\noindent\textbf{Standalone PHP test scripts.}
To evaluate the basic functionality of the translated PHP interpreter, we collected 1,143 PHP scripts for testing.
Among them, 276 are PHP engine tests shipped with PHP~5.0.0\footnote{We collected 281 shipped PHP scripts that test features within our translation scope. We filter out 5 of them: 2 on which the C engine triggers a memory bug, and 3 that do not apply to PHP~5.0.0 or to the environment we run them in.},
867 created by us for additional functionality measurement.
These tests constitute 81\% line coverage of the Zend engine code (\texttt{Zend/} folder) in total.
For each test PHP script, we compare the execution result of the Rust engine to the C engine,
and mark as pass only when the outputs are byte-identical between the two engines.
Under this standard, our Rust translation behaves identically to the C engine on all 1,143 PHP scripts we tested.

\begin{table*}[t]
\centering
\footnotesize
\setlength{\tabcolsep}{3.5pt}
\begin{threeparttable}

\caption{The eight applications and their results. PEAR is a PHP package manager and base library. \emph{Entry points} are PHP scripts that a browser can request directly, such as \texttt{index.php}.}
\label{tab:app-subjects}

\begin{tabular*}{\textwidth}{@{\extracolsep{\fill}}l l l r l r r@{}}
\toprule

Application & Version & Type & \makecell{Lines of \\ PHP code} & Dependencies & \makecell{Covered\\entry points} & \makecell{Identical\\Requests} \\
\midrule
Coppermine~\cite{coppermine1427}      & 1.4.27 & gallery & 124{,}522 & MySQL 4.1.15                   & 57/62 (92\%)   & 111/111 \\
Gallery~\cite{gallery1510}         & 1.5.10 & gallery & 41{,}086  & --                             & 80/110 (73\%)  & 76/76   \\
Nucleus CMS~\cite{nucleus324}     & 3.24   & blog    & 18{,}010  & MySQL 4.1.15, libxml2 2.9.14 (C engine only)   & 11/12 (92\%)   & 27/27   \\
phpBB~\cite{phpbb2022}           & 2.0.22 & forum   & 29{,}685  & MySQL 4.1.15                   & 24/45 (53\%)   & 52/52   \\
Simple PHP Blog~\cite{sphpblog084} & 0.8.4  & blog    & 14{,}316  & --                             & 63/72 (88\%)   & 71/71   \\
WordPress~\cite{wordpress152}       & 1.5.2  & blog    & 23{,}525  & MySQL 4.1.15, expat 2.6.1 (C engine only)      & 51/56 (91\%)   & 124/124 \\
WordPress~\cite{wordpress2.0.11}       & 2.0.11 & blog    & 30{,}073  & MySQL 4.1.15, expat 2.6.1 (C engine only)      & 49/56 (88\%)   & 122/122 \\
PEAR~\cite{pear136}            & 1.3.6  & library & 9{,}443   & expat 2.6.1 (C engine only)                    & --             & 2/2     \\
\midrule
Total           &        &         &  &                                & 335/413 (81\%) & 585/585 \\
\bottomrule
\end{tabular*}
\end{threeparttable}
\end{table*}

\noindent\textbf{Real-world PHP applications.}
Individual PHP tests check a few language features at a time.
PHP is a server-side language that powers web applications that run thousands of language features.
To further understand whether our rewrite is compatible with these applications,
we additionally select \textbf{seven} popular PHP applications plus \textbf{one} PHP library for evaluation, from the PHP~5 era and in versions that run on PHP~5.0.0, as shown in Table~\ref{tab:app-subjects}.

A PHP application normally contains multiple source PHP scripts.
When a browser asks for a page, for example \texttt{/index.php?p=3}
the web server starts the PHP interpreter on the script \texttt{index.php}.
The script reads the request input, such as the parameters, form fields, and the cookies.
It may then read or write the database.
Finally, it prints the response, which includes a few header lines followed by the page itself, usually HTML.

We build C PHP~5.0.0 from its original C source with the PHP extensions these applications need: the MySQL client, sessions, regular expressions (PCRE), compression (zlib), and XML parsing.
Our translation scope covers these necessary extensions (some of them use Rust crates; see Appendix~\ref{app:impl}), and therefore the safe Rust PHP calls them \emph{natively} without unsafe code or FFI.
We use MySQL~4.1.15~\cite{mysql} as the database and Apache~2.4.58~\cite{apache} as the web server.

The applications use consistent storage, which we call a \emph{fixture}, including database records and application settings that influence PHP execution results.
Therefore, the comparison should be under the same fixture.
We record this storage right after the application is initialized,
and use this fixture for every request.
We compare the output HTML between two engines byte by byte,
and mark it as a pass when the two sides are identical.
The page comparison ignores nondeterministic sections such as times and session identifiers.

{\bf Result III:} In total, our safe Rust translation behaves equivalently to the C PHP engine on 585 of 585 requests.
We run a further test for WordPress by adding content to the database,
to simulate a real usage scenario where the website contains blogs and comments.
All 78 requests we test remain byte-identical under real content for WordPress. More details are in Appendix~\ref{app:impl}.

\begin{tcolorbox}[
  colback=blue!4,
  colframe=blue!35!black,
  boxrule=0.5pt,
  arc=1pt,
  left=6pt,
  right=6pt,
  top=5pt,
  bottom=5pt
]

Our approach improves security while preserving functionality: the translated Rust engine successfully runs existing PHP tests and several real-world applications.
\end{tcolorbox}

%% file: sections/10-case-study.tex
\section{Case Study on a Recent PHP Vulnerability}
\label{sec:case-study}

Recently, two PHP vulnerabilities have been reported that can lead to shell execution~\cite{php2shell}.
One of them is triggered by a native memory use-after-free bug in a PHP feature that occurs as early as PHP~5.4.0.
To test whether our translation techniques can protect against this vulnerability,
we extend our safe Rust translation to cover the vulnerable feature in PHP~5.4.0.
This extended translation behaves safely on the same trigger and executes normally.
Appendix~\ref{app:case-study} explains more details.

%% file: sections/11-related.tex
\section{Related Work}
\label{sec:related}

Our work builds on research in C-to-Rust translation, memory safety, and formal verification of language interpreters.
While prior work has made progress in eliminating native memory violations and establishing the correctness of language implementations,
our work addresses a distinct challenge: ensuring that a native memory-safe interpreter also respects the logical memory semantics of the language.
We establish this guarantee for PHP by defining an abstract PHP memory machine and verifying that a core fragment of our safe-Rust reimplementation refines the abstract machine.

\noindent\textbf{C to Rust translation.}
Translating legacy C software into Rust has emerged as a promising approach to eliminating native memory-safety vulnerabilities.
Early translation systems, such as C2Rust, primarily preserve the structure and low-level semantics of C programs,
often relying on unsafe Rust to accommodate raw pointers and unrestricted aliasing~\cite{c2rust, corrode}.
Subsequent approaches use ownership inference, program analysis, and source transformations to reduce unsafe code or generate safe Rust directly~\cite{emre2021,hong2024tag}.
More recently, LLM-based translation techniques and coding agents have improved the feasibility of translating larger and more complex systems~\cite{actor,wang2026reboot, nitin2025c2saferrust}.

These approaches primarily evaluate translation success through compilation, native memory safety, and functional compatibility.
However, eliminating unsafe memory operations does not necessarily eliminate logical memory defects inherited from the original implementation.
Our work addresses this gap by specifying and verifying that the permitted memory-state effects of PHP operations respect additional safety constraints.
Thus, rather than relying on safe Rust as the sole safety improvement, we establish an additional, interpreter-specific safety guarantee.

\noindent\textbf{Memory safety and Rust guarantees.} Memory safety has traditionally been characterized in terms of spatial and temporal validity of memory accesses, including the prevention of out-of-bounds accesses and use-after-free~\cite{nagarakatte2024memory}.
Beyond these operational characterizations, Azevedo de Amorim et al.~\cite{meaning-of-memory-safety} provide a semantic account of memory safety based on noninterference: a computation should neither affect nor be affected by memory outside its reachable state.
Rust enforces native memory safety through ownership, borrowing, and runtime checks, while formal developments such as RustBelt establish semantic soundness results for a realistic subset of Rust~\cite{rustbelt}.
Recent work has begun to examine memory-safety requirements beyond the validity of native memory accesses at the language level.
For example, CMASan addresses memory safety in custom memory allocators~\cite{hong2025cmasan},
while a recent analysis of AI-generated memory-safety proofs~\cite{anthropic2026formalverification, mathur2026verified} highlights how seemingly strong guarantees can overlook violations of logical object boundaries and lifetimes.

Our work considers a different abstraction boundary.
The Rust language protects the native memory used to implement the interpreter, but it does not, by itself, guarantee the logical memory safety of the PHP interpreter.
Our approach protects logical memory safety beyond native memory safety.

\noindent\textbf{Formal verification of language interpreters.}
Prior work has developed mechanized language semantics and used them to establish correctness properties of reference interpreters and language implementation mechanisms.
For JavaScript, JSCert~\cite{bodin2014trusted} provides a mechanized specification of ECMAScript in Coq,
while JSRef is an executable reference interpreter proved correct with respect to that specification.
For PHP, KPHP~\cite{filaretti2014executable} develops an executable operational semantics for a substantial subset of the language, enabling formal reasoning about PHP programs and demonstrating program verification through model checking.
These works focus on specifying the language semantics, %
but they do not by themselves establish safety properties of a real-world interpreter implementation.

Tozawa et al.~\cite{tozawa2009copy} show that PHP's copy-on-write optimization
can violate the language's copy semantics without any invalid memory access,
and design a corrected scheme.
Their work is complementary to ours: they focus on designing a correct CoW
mechanism for PHP, whereas we focus on specifying the interpreter's entire memory
behavior as an abstract PHP memory machine.
We then prove that the real implementation refines this machine, so that the
guarantees established at the abstract level carry over to the implementation.

%% file: sections/12-conclusion.tex
\section{Conclusion}
\label{sec:conclusion}

Rewriting programs in safe Rust eliminates native memory-safety violations, but does not necessarily prevent security-relevant logical errors in managing the program's state.
We propose logical memory safety, an additional layer of guarantees beyond native memory safety.
We instantiate this guarantee for PHP by formally defining the abstract PHP machine.
We translate PHP~5.0.0 to fully safe Rust, and establish these properties using refinement-based verification for a core fragment of our implementation.
Together, we address both native and logical memory violations in our translated PHP interpreter.

%% file: sections/99-appendix.tex
\appendices

\section{Ethical Considerations}
\label{sec:ethics}

This work investigates improving the memory safety of the existing C PHP interpreter by translating it into safe Rust. Beyond native memory safety, we propose logical memory safety and formally verify the resulting Rust translation against this property.

Our evaluation uses previously known vulnerabilities and does not report newly discovered vulnerabilities. This research does not involve developing new exploitation techniques or attacking deployed systems. We use these vulnerability cases to evaluate the effectiveness of our proposed safety property and Rust translation.

The proposed approach is defensive in nature. However, incorrect translations or incomplete verification assumptions could leave security issues unresolved. We therefore distinguish the guarantees established by our formal verification from properties that remain outside its scope.

\input{sections/appendix/impl}

\input{sections/appendix/security}
\input{sections/appendix/model}
\input{sections/appendix/case-study}

%% file: sections/appendix/impl.tex
\section{Implementation and Evaluation Details}
\label{app:impl}

\subsection{Translation scope}
\label{app:impl-scope}

Table~\ref{tab:scope} lists each part of PHP~5.0.0 that we translate,
with its size in C and in safe Rust.
Besides the main engine, which is about 37k lines of code in C, our translation also covers PHP libraries and extensions.
In total, the translation covers roughly 115{,}000 lines of C and has
77{,}000 lines of safe Rust.

A module is \emph{full} when the translation covers all of it, and
\emph{partial} when the rewrite finishes only the part our evaluation needs.
A module is \emph{reimplemented} when its C code is a thin layer over an
external C library.
Instead of calling from FFI, the translation rewrites these libraries in safe Rust, through crates
for PCRE and zlib, and through manual implementation of the MySQL
client protocol for MySQL.
For each extension, the last column gives how many of the PHP functions
it provides are covered in the translation.

\begin{table}[t]
\centering
\footnotesize
\setlength{\tabcolsep}{4pt}
\caption{Scope of the translation. Lines of code are non-blank,
non-comment lines.}
\label{tab:scope}
\begin{tabular}{@{}l r r l r@{}}
\toprule
Module & C & Rust & Status & Functions \\
\midrule
\multicolumn{5}{@{}l}{\emph{Engine}} \\
Zend/               & 37{,}750 & 35{,}012 & full          & --        \\
\addlinespace[2pt]
\multicolumn{5}{@{}l}{\emph{Runtime}} \\
main/               & 17{,}499 & 4{,}869  & partial       & --        \\
sapi/ (CLI and CGI) & 2{,}285  & 975      & partial       & --        \\
\addlinespace[2pt]
\multicolumn{5}{@{}l}{\emph{Standard library and extensions}} \\
ext/standard        & 36{,}656 & 23{,}309 & partial       & 448/498 \\
ext/pcre            & 1{,}203  & 1{,}836  & reimplemented & 7/7     \\
ext/mbstring        & 7{,}488  & 2{,}753  & partial       & 12/60   \\
ext/exif            & 3{,}244  & 1{,}531  & partial       & 2/5     \\
ext/session         & 2{,}249  & 1{,}240  & partial       & 16/20   \\
ext/mysql           & 2{,}017  & 2{,}242  & reimplemented & 60/60   \\
ext/xml             & 1{,}898  & 2{,}667  & full          & 24/24   \\
ext/calendar        & 1{,}255  & 362      & partial       & 1/18    \\
ext/zlib            & 869      & 119      & reimplemented & 20/22   \\
ext/bcmath          & 469      & 468      & full          & 10/10   \\
ext/ctype           & 147      & 87       & full          & 11/11   \\
\midrule
Total               & 115{,}029 & 77{,}470 &              &         \\
\bottomrule
\end{tabular}
\end{table}

\subsection{Dependencies}
\label{app:impl-deps}

Table~\ref{tab:crates} lists the crates the translation depends on: six
directly, and twelve in all.
Apart from these crates, the Rust translation contains no unsafe code.
We enforce this with \texttt{\#![forbid(unsafe\_code)]} on our Rust translation.

\begin{table}[t]
\centering
\footnotesize
\setlength{\tabcolsep}{3pt}
\caption{Crates used by the Rust translation. Downloads are the total
number of downloads (in millions) within 90 days.}
\label{tab:crates}
\begin{tabular*}{\columnwidth}{
    @{\extracolsep{\fill}} l l l r @{}
}
\toprule
Crate & Version & Used for & Downloads \\
\midrule
\multicolumn{4}{@{}l}{\emph{Direct dependencies}} \\
slotmap        & 1.1.1  & PHP references         & 39.8M  \\
regex          & 1.12.3 & PCRE                   & 331.5M \\
regex-automata & 0.4.16 & PCRE                   & 363.0M \\
fancy-regex    & 0.19.0 & PCRE                   & 106.2M \\
memchr         & 2.8.0  & Substring search       & 469.9M \\
miniz\_oxide   & 0.9.1  & zlib                   & 293.3M \\
\addlinespace[2pt]
\multicolumn{4}{@{}l}{\emph{Indirect dependencies}} \\
aho-corasick   & 1.1.4  & regex, regex-automata  & 346.4M \\
regex-syntax   & 0.8.10 & regex, regex-automata  & 373.7M \\
bit-set        & 0.8.0  & fancy-regex            & 172.7M \\
bit-vec        & 0.8.0  & bit-set                & 201.4M \\
adler2         & 2.0.1  & miniz\_oxide           & 218.3M \\
version\_check & 0.9.5  & slotmap (build time)   & 308.4M \\
\bottomrule
\end{tabular*}
\end{table}

\subsection{Build configuration}
\label{app:impl-build}

Table~\ref{tab:builds} lists the build used for each evaluation.
All C builds use GCC~13.3.0 and also pass \texttt{-std=gnu89} and
\texttt{-fcommon}, \texttt{-falign-functions=64}.
All Rust builds use rustc~1.97.1 in release mode with
\texttt{lto = "fat"}, \texttt{codegen-units = 1}, \texttt{-align-all-functions=6},
\texttt{overflow-checks = true} and an inline threshold of 1000.

\begin{table}[t]
\centering
\footnotesize
\setlength{\tabcolsep}{3pt}
\caption{The compilation flag used for C PHP~5.0.0 and the Rust translation.}
\label{tab:builds}
\begin{tabular}{@{}l >{\raggedright\arraybackslash}p{2.1cm}
                    >{\raggedright\arraybackslash}p{4.75cm}@{}}
\toprule
Engine & Used for & Build \\
\midrule
C    & Safety (Section~\ref{sec:eval-security})
     & \texttt{-O0}, AddressSanitizer\\
C    & Bug confirmation (Appendix~\ref{app:sec-bench})
     & \texttt{-O0 -g}, AddressSanitizer, the engine's memory cache off
       and allocation sizes not rounded \\
C    & Performance
     & \texttt{-O3 -march=native -mtune=native}
       \texttt{-fomit-frame-pointer -flto}, plus the extensions for the functionality tests that need them \\
C    & Functionality
     & \texttt{-O0}, plus MySQL, sessions, PCRE, zlib and XML for testing applications \\
\addlinespace[2pt]
Rust & Safety, performance, functionality
     & \texttt{cargo build --release --features fast-paths} \\
\bottomrule
\end{tabular}
\end{table}

\texttt{fast-paths} switches 4 slow verified features to optimized implementations, which Verus does not support.
We verified the refactored, slower version of the same code, while using the fast version for evaluation.
We manually checked that each fast version is semantically equivalent to its verified
version.
Together, the four fast paths are fewer than 60 lines of Rust code edits.

\subsection{Evaluate on WordPress with Real Content.}

\begin{table}[t]
\centering
\small
\begin{threeparttable}
\caption{WordPress over real content.}
\label{tab:app-wp-content}
\begin{tabular}{@{}l l r r r@{}}
\toprule
Version & Data set & Posts & Comments & Identical Requests \\
\midrule
1.5.2  & wptest & 52  & 31    & 15/15 \\
       & wptt   & 79  & 34    & 15/15 \\
       & books  & 601 & 1,201 & 15/15 \\
\midrule
2.0.11 & wptest & 53  & 31    & 11/11 \\
       & wptt   & 80  & 34    & 11/11 \\
       & books  & 602 & 1,201 & 11/11 \\
\bottomrule
\end{tabular}
\end{threeparttable}
\end{table}

We run a further test for WordPress by adding content to the database,
to simulate a real usage scenario where the website contains blogs and comments.
We load three real data sets:
two standard WordPress test-data exports~\cite{wordpressThemeTestData,wptest},
and a set built from five Project Gutenberg books~\cite{gutenbergCorpus}.
When testing with these real contents, the responses remain identical between the C and Rust engines on all 78 requests we tested,
as shown in Table~\ref{tab:app-wp-content}.

%% file: sections/appendix/security.tex
\section{Security Corpus}
\label{app:sec-bench}

\subsection{How we collect the bugs}
\label{app:sec-collect}

The corpus has 173 memory bugs in PHP~5.0.0.
We collected them from the Git commit history of PHP.
We manually reviewed and filtered out duplicate cases with the same root cause.
We confirmed each bug in PHP~5.0.0 with several kinds of evidence.
The 142 native memory bugs are confirmed with AddressSanitizer.
We compile PHP~5.0.0 with the flags \texttt{-O0}, \texttt{-g},
\texttt{-fno-omit-frame-pointer}, \texttt{-fsanitize=address} and
\texttt{-fno-sanitize-address-use-after-scope}, and we turn off the
engine's cache of freed memory and its rounding of allocation sizes.
The 31 PHP logical semantic errors do not corrupt memory, so we rely on
three sources of evidence:
\begin{enumerate}
 \item We run the trigger on different PHP versions and compare
       the results. A different result suggests that the behavior is
       unstable.
 \item When a fix commit exists in the Git history, its message gives
       further evidence that the bug is real.
 \item Finally, we inspect each bug by hand and confirm that it breaks a rule of
       PHP's memory semantics, such as a wrong reference count.
\end{enumerate}

All of these bugs were fixed during PHP's history, and none is present
in PHP today.
For 147 of them, each is mentioned explicitly in a fix commit in the history.
The other 26 were fixed indirectly: their faulty code disappeared because some commits restructured the implementation.

\subsection{The groups}
\label{app:sec-groups}

Table~\ref{tab:sec-bench} in Section~\ref{sec:bug-landscape} sorts the
173 bugs by defect class.
A bug belongs to the library when its faulty line is in the standard library or in an extension, and to the engine otherwise.
142 of them are C raw memory corruption. We omit their explanation due to space limits.

31 bugs are PHP memory semantic errors.
These bugs do not corrupt the native memory.
They violate the intended semantics of PHP.
\textbf{(1) Extra reference binding (11):} Two variables or elements
become one reference although the program never bound them with
\texttt{\&}, so a write through one changes the other.
\textbf{(2) Lost reference binding (8):} The program bound two names
with \texttt{\&}, but the engine gives one of them new storage instead
of writing through the binding, so later writes no longer reach the
other.
\textbf{(3) Improper resource lifetime (5):} The engine gets the
sharing of a value wrong.
A write through one holder is then seen by another.
\textbf{(4) Missing memory release (5):} The engine never releases
a value.
\textbf{(5) Incorrect reference count (1):} The engine adds one to
an object's reference count and never removes it.
\textbf{(6) Premature release (1):} Deleting one array element
destroys a different element that is still in use.

%% file: sections/appendix/model.tex
\section{The Abstract PHP Machine}
\label{app:model}

\subsection{Holder counts and fresh names}
\label{app:holder-counts}

The holders of a VStore $v$ are the VSlots that point to it, and the
holders of an HStore $h$ are the VStores that hold its handle:
\begin{equation*}
\begin{aligned}
 \mathrm{rc}_s(v) &= |\{\ell\in\mathsf{Slot}_s\mid \mathrm{pt}_s(\ell)=v\}|,\\
 \mathrm{rc}_s(h) &= |\{v\in\mathsf{Store}_s\mid \mathrm{val}_s(v)=h\}|.
\end{aligned}
\end{equation*}

\subsection{The copy}
\label{app:copy}

Value assignment copies the content $c$ of a VStore with
$\mathrm{copy}_s(c)=(c',\Delta)$, where $c'$ is the copied content and
$\Delta$ holds the entries of the storage that the copy creates.
A scalar and an object handle are copied as they are, and an array is
duplicated:

\begin{equation*}
\begin{aligned}
 \mathrm{copy}_s(c) &= (c,\emptyset)
   &&\text{if } c\in\mathsf{Scalar}\text{ or }\mathrm{kind}_s(c)=\mathsf{obj},\\
 \mathrm{copy}_s(c) &= \mathrm{dup}_s(c)
   &&\text{if } \mathrm{kind}_s(c)=\mathsf{arr},
\end{aligned}
\end{equation*}
\begin{equation*}
 \mathrm{dup}_s(h) = \Bigl(h',\ \{h'\mapsto\mathrm{kind}_s(h)\}
   \cup\textstyle\bigcup_{h.k\in\mathsf{Slot}_s}\Delta_k\Bigr),
\end{equation*}
where, for each element $h.k$ of $h$,
\begin{equation*}
\Delta_k=
\begin{cases}
 \{h'.k\mapsto u_k\}
   & \text{if } \mathrm{rc}_s(u_k)>1,\\
 \{h'.k\mapsto v_k,\ v_k\mapsto c_k\}\cup\Delta'_k
   & \text{otherwise,}
\end{cases}
\end{equation*}
with $u_k=\mathrm{pt}_s(h.k)$ and $(c_k,\Delta'_k)=\mathrm{copy}_s(\mathrm{val}_s(u_k))$.
The names $h'$ and $v_k$, and those of the nested copies, are fresh in $s$
and all distinct.

So the new HStore $h'$ has the keys of $h$.
An element that is a reference, that is, whose VStore has more than one
holder, stays shared: the new element points to the same VStore.
Every other element gets a new VStore with a copy of its content.
A nested array is therefore duplicated in turn, to any depth, and a nested
object is not, because only its handle is copied.
Cloning an object applies $\mathrm{dup}_s$ to the object itself.
Verus checks that this recursion ends for every copy that a rule makes.

\subsection{The transition rules}
\label{app:transition-rules}

The seven operations have eight rules, because creating a variable and
creating an element are separate rules.
In the rules, $\ell_p=\llbracket p\rrbracket_s$,
$\ell_q=\llbracket q\rrbracket_s$, $\ell_i=\llbracket p_i\rrbracket_s$,
$v_p=\mathrm{pt}_s(\ell_p)$ and $v_q=\mathrm{pt}_s(\ell_q)$.
A rule applies only when the VSlots it uses exist.
\begin{gather*}
\textsc{Create-Var}\quad
\frac{x\notin\mathsf{Slot}_s\qquad w\notin\mathsf{Store}_s}
     {s\xrightarrow{\mathsf{create}\ x}s[x\mapsto w,\ w\mapsto\mathsf{null}]}
\\[8pt]
\textsc{Create-Elem}\quad
\frac{\begin{array}{c}
       \mathrm{val}_s(v_p)=h\in\mathsf{Heap}\\
       h.k\notin\mathsf{Slot}_s\qquad w\notin\mathsf{Store}_s
      \end{array}}
     {s\xrightarrow{\mathsf{create}\ p[k]}s[h.k\mapsto w,\ w\mapsto\mathsf{null}]}
\\[8pt]
\textsc{New}\quad
\frac{h\notin\mathsf{Heap}_s}
     {s\xrightarrow{p:=\mathsf{new}\ \kappa}s[h\mapsto\kappa,\ v_p\mapsto h]}
\\[8pt]
\textsc{Assign}\quad
\frac{(c,\Delta)=\mathrm{copy}_s(\mathrm{val}_s(v_q))}
     {s\xrightarrow{p:=q}(s\uplus\Delta)[v_p\mapsto c]}
\\[8pt]
\textsc{Bind}\quad
\frac{}
     {s\xrightarrow{p=\&q}s[\ell_p\mapsto v_q]}
\\[8pt]
\textsc{Clone}\quad
\frac{\begin{array}{c}
       \mathrm{val}_s(v_q)=h\qquad \mathrm{kind}_s(h)=\mathsf{obj}\\
       (h',\Delta)=\mathrm{dup}_s(h)
      \end{array}}
     {s\xrightarrow{p:=\mathsf{clone}\ q}(s\uplus\Delta)[v_p\mapsto h']}
\\[8pt]
\textsc{Unset}\quad
\frac{}
     {s\xrightarrow{\mathsf{unset}\ p}s\setminus\ell_p}
\\[8pt]
\textsc{Compute}\quad
\frac{d=\mathcal I(o)(\mathrm{view}_s(\ell_1,\ldots,\ell_n))}
     {s\xrightarrow{p:=\mathsf{compute}(p_1,\ldots,p_n)\ \mathsf{by}\ o}s[v_p\mapsto d]}
\end{gather*}
In words:
\begin{itemize}
 \item \textsc{Create-Var} and \textsc{Create-Elem}: a new VSlot, either a
       variable $x$ or the element $h.k$ of an array or object, comes into
       being and points to a new VStore that holds \texttt{null}.
 \item \textsc{New}: the VStore of $p$ receives the handle of a new, empty
       HStore of kind $\kappa$.
 \item \textsc{Assign}: the content of the VStore of $q$ is copied into the
       VStore of $p$. No VSlot changes where it points.
 \item \textsc{Bind}: the VSlot of $p$ now points to the VStore of $q$.
 \item \textsc{Clone}: the VStore of $p$ receives the handle of a copy of
       the object of $q$.
 \item \textsc{Unset}: the VSlot of $p$ disappears. Its VStore stays, even
       when nothing holds it any more.
 \item \textsc{Compute}: the VStore of $p$ receives the scalar that
       $\mathcal I(o)$ gives for the view of the sources.
\end{itemize}

\subsection{The VSlots an operation names}
\label{app:operand-slots}

Definition~\ref{def:frame-property} uses $X_s(a)$, the VSlots of the operands of $a$, in order:
\begin{itemize}
 \item $(x)$ for $\mathsf{create}\ x$, and $(h.k)$ for
       $\mathsf{create}\ p[k]$ with $h=\mathrm{val}_s(v_p)$, that is,
       the VSlot that the operation creates;
 \item $(\ell_p)$ for $p:=\mathsf{new}\ \kappa$ and for $\mathsf{unset}\ p$;
 \item $(\ell_p,\ell_q)$ for $p:=q$, $p=\&q$ and $p:=\mathsf{clone}\ q$;
 \item $(\ell_p,\ell_1,\ldots,\ell_n)$ for
       $p:=\mathsf{compute}(p_1,\ldots,p_n)\ \mathsf{by}\ o$.
\end{itemize}

\subsection{The links of a copy}
\label{app:copy-connections}

Value assignment and cloning write a content $c$ into the VStore $v$ of
$p$ and add the entries $\Delta$ of a copy:
$(c,\Delta)=\mathrm{copy}_s(\mathrm{val}_s(v'))$ for $p:=q$, and
$(c,\Delta)=\mathrm{dup}_s(h)$ for $p:=\mathsf{clone}\ q$, where $h$ is
the object of $q$.
$\mathrm{CopyConn}_s(a)$ holds the link from $v$ to $c$ when $c$ is a
handle, and every link that starts in $\Delta$:
\begin{equation*}
\begin{aligned}
 \mathrm{CopyConn}_s(a)={}&\{(v,c)\mid c\in\mathsf{Heap}\}\\
  &\cup\{(\ell'',w)\mid \ell''\mapsto w\in\Delta\}\\
  &\cup\{(w,h'')\mid w\mapsto h''\in\Delta,h''\in\mathsf{Heap}\}.
\end{aligned}
\end{equation*}
The first link ends at the shared object or at the new array.
A link from a new element may end at an old VStore, which keeps a
reference shared, and a link from a new VStore may end at an old object,
which keeps an object handle shared.

\subsection{Renaming names}
\label{app:renaming}

A renaming $\rho$ is a one-to-one map from VStore names onto VStore names
and from HStore names onto HStore names.
It renames an element slot $h.k$ to $\rho(h).k$ and a handle $h$ to
$\rho(h)$, and it leaves variables, keys, scalars, kinds and holder counts
unchanged.
$\rho(X)$ renames each VSlot of a sequence $X$.
$\rho(f)$ renames every name in a fragment $f$: an entry $\ell\mapsto v$
becomes $\rho(\ell)\mapsto\rho(v)$, an entry $v\mapsto c$ becomes
$\rho(v)\mapsto\rho(c)$, and an entry $h\mapsto\kappa$ becomes
$\rho(h)\mapsto\kappa$.
A view is renamed in the same way.

Using the notation of Definition~\ref{def:frame-property}, let
$f=s'\upharpoonright_{N\cup N'}$ and
$g=t'\upharpoonright_{M\cup M'}$.
Then $g\cong_\rho f$ holds when some one-to-one renaming $\sigma$ maps $f$
onto $g$, counts included, and
\begin{itemize}
 \item $\sigma$ agrees with $\rho$ on every name of $N$;
 \item every name of $f$ that exists in $s$ is mapped as $\rho$ maps it,
       and its image exists in $t$;
 \item every name of $f$ that does not exist in $s$ is mapped to a name
       that does not exist in $t$.
\end{itemize}
So $\sigma$ follows $\rho$ on old storage, and it may choose freely only
for the storage that the two transitions create.

\subsection{Abstract-machine safety}
\label{app:safety-proof}

\begin{proof}[Proof sketch of Theorem~\ref{thm:abstract-machine-safety}]
It suffices to show that every transition satisfies the frame property
and sharing integrity. Let $s\xrightarrow{a}s'$ be a transition, and let
$N,N'$ be its name sets from Definition~\ref{def:frame-property}.
We proceed by cases on the transition rule.

For read confinement, consider two transitions satisfying the premise of
Definition~\ref{def:frame-property}. Their operand slots, contents, and
holder counts correspond under $\rho$. For assignment and cloning,
induction on the copy rules gives corresponding results and fresh
allocations: the holder counts determine whether an element remains
shared or its contents are copied. Computation returns the same scalar
because $\mathcal I$ depends only on the ordered operand view and is
invariant under renaming. The remaining rules perform corresponding map
updates. The original operand slots therefore reach corresponding
storage afterward. Since the initial holder counts agree, corresponding
changes to connections also give matching final counts. Matching the
fresh allocations then gives the required congruence.

For write confinement, each rule changes existing entries only at the
destination slot or its VStore, both of which belong to $N$. Fresh
storage is reachable from the created slot or updated destination in
$s'$, and hence belongs to $N'$. For assignment and cloning, this follows
by induction on the copy rules. Holder counts change only at the
targets of added or removed connections, which also belong to
$N\cup N'$. The state outside $N\cup N'$ therefore remains unchanged.

Finally, sharing integrity follows by checking the connection changes
against Table~\ref{tab:allowed-connections}. Each rule removes the
connections specified by $\mathrm{Rm}_s(a)$ and adds only connections
permitted by $\mathrm{Add}_s(a)$. In particular, assignment and cloning
introduce the destination link, if any, and the links of the returned copy, as
specified by $\mathrm{CopyConn}_s(a)$. Every transition thus satisfies
both required properties, and the theorem follows from
Definition~\ref{def:logical-memory-safety}.
\end{proof}

%% file: sections/appendix/case-study.tex
\section{Case Study on Recent PHP Vulnerabilities}
\label{app:case-study}

The use-after-free comes from a global variable that outlives the memory it points to.
In a long-running PHP server, the session module keeps the current session ID in \texttt{PS(id)}, a global pointer that refers to memory owned by a single request.
At the end of the request, the shutdown procedure frees the request's memory, including the session ID, but does not clear the global pointer.
The pointer is therefore left dangling.
The next request can then read this dangling pointer through upload progress, a feature added in PHP~5.4.0, and trigger the use-after-free.

To test our translation against the same vulnerability, we added the 5.4.0 upload-progress feature to our rewrite in 276 lines of safe Rust.
On the same trigger path (two consecutive requests), the translation runs safely, with no native memory corruption, which safe Rust eliminates by construction.
Therefore, the remote code execution enabled by the raw memory bug will be stopped.

Applying our proof strategy to the newly backported feature can prevent the logical form of this violation (reads into a dangling value in the logical program state) as well.
The proof obligation requires the program state to be well-defined, so a dangling value in the state cannot pass the proof.